\documentclass[twocolumn,american]{revtex4-2}
\usepackage[T1]{fontenc}
\usepackage[utf8]{inputenc}
\usepackage{amsmath}
\usepackage{amssymb}
\usepackage{graphicx}
\usepackage{esint}
\usepackage{babel}
\begin{document}
\title{Quasiconformal geometry and elastic anisotropy in liquid crystals}
\author{T. J. Atherton}
\email{timothy.atherton@tufts.edu}

\affiliation{Tufts University Department of Physics \& Astronomy, 574 Boston Ave,
Medford, MA 02155, USA.}
\begin{abstract}
Unequal splay and bend elastic constants endow planar nematic liquid
crystals with a director-dependent anisotropic geometry. Here, we
show that this geometry is quasiconformal: the magnitude of the associated
Beltrami coefficient is fixed by the elastic constants, while its
phase depends on the director relative to a compatible rotating frame
introducing an $SO(2)$ spin connection. We construct explicit maps
for a hybrid-aligned film and isolated defect textures, generalizing
the conformal construction of isotropic defects. These results provide
a new insight into the classical Dzyaloshinskii defect solutions,
which are shown to define an auxiliary reference cone on which the
defect texture is quasiconformally distorted before being projected
conformally into physical space. Numerical simulations of constrained
$+2$ defects reveal core morphologies consistent with the quasiconformal
geometry. Our results suggest that quasiconformal maps provide a broader
geometric language for anisotropy in liquid crystals and other materials.
\end{abstract}
\maketitle

\section{Introduction}

Many materials exhibit elastic anisotropy, where the energetic cost
of a deformation depends on the direction in which it is applied.
Anisotropy can greatly affect the structure and mechanical response
of the material, shaping its equilibrium structure or inducing multiple
stable states. In crystalline solids, anisotropy is constrained by
the lattice symmetry and is therefore orientationally fixed at least
within a domain or grain. 

In a liquid crystal, however, the local order and any dependent anisotropic
material properties are free to vary spatially. The Frank-Oseen free
energy for nematic liquid crystals, for example, distinguishes splay,
twist and bend deformations with associated elastic constants $K_{1}$,
$K_{2}$ and $K_{3}$. Differences between these constants can drive
the formation of modulated phases\citep{Selinger2022}, promote spontaneous
chiral symmetry breaking in achiral liquid crystals\citep{Jeong2014}
and affect the structure adopted, e.g. the shape of topological defect
cores\citep{Schimming2020} or the morphology of LC droplets \citep{DeBenedictisAtherton2016,SchimmingVinals2022}.
Recent work on active nematics shows dynamical consequences of elastic
anisotropy\citep{Houston2026}.

Anisotropy can also be used to control the structure, e.g. by driving
topological transformations between tactoidal and toroidal droplets\citep{Koizumi2023}.
Competition between elastic distortions can also be exploited to control
structure through surface patterning, inducing preferred director
orientations \citep{AthertonSambles2006} and enabling arbitrary alignment
directions through the design of patterned substrates \citep{AnquetilDeck2012,AnquetilDeck2013,DeBenedictis2015}. 

While anisotropy leads to a wide variety of physical effects, a very
common approximation is to treat the elastic constants as equal, with
a considerable payoff in mathematical simplicity: stationary solutions
of the Frank-Oseen energy for planar nematics satisfy Laplace's equation
away from defects, perhaps the most studied partial differential equation
known. In two dimensions, there is a natural connection with complex
analysis, and a specific class of complex analytic functions called
conformal mappings provide a powerful explicit method for analytically
constructing liquid crystal textures\citep{ChandlerSpagnolie2023,ChandlerSpagnolie2024Interactions,Chandler2025}.
Such mappings can be interpreted geometrically: they describe conformal
transformations that locally preserve angles while allowing spatially
varying dilation and rotation \citep{needham2023visual}. When the
elastic constants are unequal, however, the Euler-Lagrange equations
are not harmonic and conformal mappings alone are insufficient to
account for the anisotropic elastic response.

In this work, we show that in the anisotropic case, conformal maps
and their associated geometry can be replaced by quasiconformal maps,
a natural generalization of conformal maps that incorporate local
anisotropic stretching while remaining grounded in complex analysis\citep{Ahlfors2006,AstalaIwaniecMartin2009}.
While mathematically important for Teichm\"uller theory, complex
dynamics and geometric function theory\citep{AhlforsBers1960,Bonk2006},
their use in physics has remained relatively scattered: they have,
for example, been used to transform anisotropic conductivity problems\citep{AstalaLassasPaivarinta2005},
design optical cloaks through transformation optics\citep{LiPendry2008}
and describe electronic dynamics in strained graphene\citep{FillionGourdeau2021}. 

A second group of applications of quasiconformal maps includes biological
morphology, where they have been used to characterize morphological
differences between species due to evolution\citep{JonesMahadevan2013}
and, recently, to describe growth and shape change of thin biological
tissues in morphogenesis\citep{CisloPavlopoulosShraiman2025}. The
disparate nature of these applications suggests that quasiconformal
maps may provide a more general language for anisotropy across physical
systems than has been appreciated. Liquid crystals are a particularly
rich setting to investigate this possibility since the local elastic
anisotropy varies spatially with the orientational order parameter,
whose equilibrium configuration is itself determined by a variational
principle, and the materials support a wide variety of topologically
nontrivial structures.

Topological defects provide a particularly interesting setting in
which to explore this geometric perspective, and have recently attracted
renewed attention \citep{schimming2026structure}. An analytical solution
for unequal splay and bend elastic constants was discovered by Dzyaloshinskii
\citep{dzyaloshinskii1970theory}, its stability subsequently studied\citep{anisimov1973new}.
These solutions have been used to infer elastic anisotropy of liquid-crystalline
polymers from experimental TEM data\citep{hudson1989frank}. More
recently, Schimming and Vi\~nals \citep{Schimming2020} simulated
anisotropic $\pm\frac{1}{2}$ defects in the $Q$-tensor framework,
demonstrating how elastic anisotropy influences the shape of the core.

To show that the elastic anisotropy of planar nematics defines a quasiconformal
geometry, and develop an explicit mapping construction for anisotropic
director configurations, the paper proceeds as follows: Following
background in Sections \ref{sec:ConformalDefects} and \ref{sec:Quasiconformal-maps},
we establish the geometric framework and illustrate it using a hybrid-aligned
nematic film in Section \ref{sec:Anisotropic-elastic-geometry}. Applying
the construction to topological defects in Section \ref{sec:Anisotropic-defect-structure}
reveals a new geometric interpretation of the classical Dzyaloshinskii
solutions: the defect closure condition defines an auxiliary conical
geometry, providing a surprising connection to recent work on nematics
on physical cones\citep{VafaZhangNelson2022,ZhangNelson2022}. The
anisotropic texture itself is constructed by the composition of a
quasiconformal map and an elementary conformal map. The associated
frame rotation introduces a spin connection with nontrivial holonomy
around the defect. Finally, numerical simulations in Section \ref{sec:Numerical-Experiments}
of $+2$ defect cores reveal that their morphologies follow the family
of shapes generated by quasiconformal mapping, even when the effective
anisotropy differs from that determined by the corresponding Frank
elastic constants. We conclude by discussing the implications of our
work for measuring elastic anisotropy and studying other kinds of
anisotropy in Sections \ref{sec:Discussion} and \ref{sec:Conclusions}.

\section{Conformal maps and Liquid Crystals}\label{sec:ConformalDefects}

In two dimensions, the structure of a nematic liquid crystal is found
by minimizing the Frank free energy, 
\begin{equation}
F=\frac{1}{2}\int_{\Omega}d^{2}x\left[K_{1}\left(\nabla\cdot\mathbf{n}\right)^{2}+K_{3}\left(\nabla\times\mathbf{n}\right)^{2}\right],\label{eq:Frank2D}
\end{equation}
where $\mathbf{n}$ is a unit vector describing the local preferred
orientation, restricted to lie in-plane, and $\nabla\times\mathbf{n}$
is the two dimensional curl. If the elastic constants are equal, $K=K_{1}=K_{3}$,
Eq. (\ref{eq:Frank2D}) simplifies to,
\begin{equation}
F=\frac{K}{2}\int_{\Omega}d^{2}x\left|\nabla\theta\right|^{2},\label{eq:Frank2D1Constant}
\end{equation}
using the parametrization $\mathbf{n}=\left(\cos\theta,\sin\theta\right)$.
Stationary states of Eq. (\ref{eq:Frank2D1Constant}) are therefore
solutions of Laplace's equation $\nabla^{2}\theta=0$ away from the
defect cores.

Complex functions are a rich source of solutions to Laplace's equation
in 2D, and hence a deep connection between liquid crystal structure
and complex analysis emerges: consider a map $z(w)$ from a reference
plane $w=u+iv$ to a physical plane $z=x+iy$. If such a map satisfies
the Cauchy-Riemann conditions, $x_{u}=y_{v},y_{u}=-x_{v}$, where
subscripts denote differentiation with respect to the subscripted
coordinate, then its real and imaginary parts both satisfy Laplace's
equation and are harmonic conjugates of each other. 

Such maps also possess a geometric interpretation: The complex derivative
$z_{w}=s(w)e^{i\gamma(w)}$ encodes a local instruction to scale an
arbitrary vector in the reference plane by a factor $s=\left|z_{w}\right|$
and rotate it by $\gamma=\arg z_{w}$. Infinitesimal circles around
a point of interest are transformed into dilated circles under the
action of the map while angles are locally preserved and hence such
maps are termed \textit{conformal}. Conformal maps can be composed,
and formulae such as the Schwarz-Christoffel map\citep{driscoll2002schwarz}
exist to map canonical domains to shapes of interest such as polygons.
Physical quantities, such as forces or torques, can be calculated
directly from such maps, and hence conformal mapping has been extensively
used as an analytical tool to study nematics\citep{ChandlerSpagnolie2023,Chandler2025}.

\begin{figure}
\includegraphics[width=1\columnwidth]{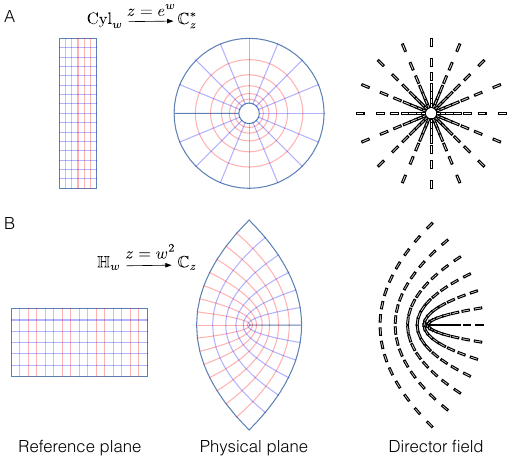}

\caption{\textbf{Conformal construction of nematic defects.} \textbf{A} a
$+1$ defect and \textbf{B} a $+\frac{1}{2}$ defect. In each case,
a spatially uniform director field in the reference domain is transformed
to the physical domain and rotated by the action of the map.}\label{fig:DefectConformal}
\end{figure}

Under a conformal map $z=f(w)$, a spatially uniform director field
transforms according to $\theta(f(w))=\arg f'(w)+\text{const}.$,
which is locally the imaginary part of the analytic function $\log f'$
and hence is harmonic wherever $f'\neq0$. Since harmonicity is preserved
by conformal maps, the resulting director satisfies $\nabla_{z}^{2}\theta=0$
away from singularities. 

As explicit examples, we construct two nematic defect structures using
conformal mapping. In Fig. \ref{fig:DefectConformal}A, we show a
map $z=e^{w}$ from a rectangle $u\in[-a,a],v\in[-\pi,\pi]$ in the
reference $w$ plane to an annulus of inner and outer radii $e^{-a}$
and $e^{a}$ in the physical $z$ plane. Identifying the upper and
lower boundaries of the rectangle, the reference domain becomes a
cylinder $\text{Cyl}_{w}$ that maps conformally to the annulus. As
$a\to\infty$, the infinite cylinder maps to the punctured complex
plane $\mathbb{C}_{z}^{*}$. Contours of $v$ in the $w$ plane become
radial spokes in the $z$ plane under the action of the map. Hence
a spatially uniform director lying along the real axis in the $w$
plane is rotated to become a $+1$ defect, $\theta(z)=\arg z$, in
the physical plane under the action of the map. 

Meanwhile in Fig. \ref{fig:DefectConformal}B, we show an analogous
construction for a $+1/2$ defect. The reference $w$-plane here is
the upper half plane $\mathbb{H}_{w}$, mapped to the physical $z$
plane $\mathcal{S}_{z}=\mathbb{C}_{z}\backslash\left[0,\infty\right)$
by the map $z=w^{2}$, where the real axis in $w$ becomes a slit
along the positive real axis in the $z$-plane. As before, a spatially
uniform director lying along the real axis is rotated to become a
$+1/2$ defect, $\theta(z)=\frac{1}{2}\arg z$ in the $z$ plane.
The director rotates by $\pi$ across the two sides of the slit, so
the nematic orientation is continuous under their identification. 

These two examples show the utility of constructing director configurations
by conformal mapping. They also provoke an obvious question: is there
an analogous construction for the anisotropic case $K_{1}\neq K_{3}$?
We will show in the remainder of the paper that quasiconformal maps
fulfill this role, and provide new insight into anisotropic elasticity
more generally.

\section{Quasiconformal maps}\label{sec:Quasiconformal-maps}

Having established conformal maps as tools to construct liquid crystal
structures, we introduce quasiconformal maps by beginning with a geometric
picture: A map from a reference $w$-plane to a physical $z$-plane
is quasiconformal if infinitesimal circles around any non-singular
point are mapped to ellipses with uniformly bounded ratios of their
principal axes. Consequently, angles are not generally preserved by
such mappings. 

A very simple example is provided by a stretch transformation. Consider
the map,

\begin{equation}
z(w)=\alpha u+iv=\frac{1}{2}\left[\left(\alpha+1\right)w+\left(\alpha-1\right)\overline{w}\right].\label{eq:stretchcomplex}
\end{equation}
where $\alpha>0$ is a real parameter, which stretches the $u$-direction
relative to the $v$-direction. As $\alpha\to1$, $z(w)\to w$ the
identity operation which is conformal. For $\alpha\neq1$, any circle
in $w$ is stretched to an ellipse with ratio of major to minor axes
$\mathcal{K}=\max(\alpha,\alpha^{-1})$; this quantity is bounded
and hence the map is quasiconformal.

Quasiconformal maps may be differentiated by defining the Wirtinger
derivatives,
\begin{align}
\frac{\partial}{\partial w} & =\frac{1}{2}\left(\frac{\partial}{\partial u}-i\frac{\partial}{\partial v}\right),\\
\frac{\partial}{\partial\bar{w}} & =\frac{1}{2}\left(\frac{\partial}{\partial u}+i\frac{\partial}{\partial v}\right),
\end{align}
so that $z_{w}=\frac{\partial z}{\partial w}$ and $z_{\bar{w}}=\frac{\partial z}{\partial\bar{w}}$;
note these are defined even for maps that are not differentiable in
the ordinary sense. Evaluating the derivative $z_{\bar{w}}$ explicitly,
\begin{equation}
z_{\bar{w}}=\frac{1}{2}\left[\left(\frac{\partial x}{\partial u}-\frac{\partial y}{\partial v}\right)+i\left(\frac{\partial x}{\partial v}+\frac{\partial y}{\partial u}\right)\right],\label{eq:WirtingerDbar}
\end{equation}
we notice the real and imaginary components correspond to the Cauchy-Riemann
equations which can be written compactly as $z_{\bar{w}}=0$. 

Since both $z_{w}$ and $z_{\bar{w}}$ are both complex numbers, they
can be related to one another by a complex number $\mu$,
\begin{equation}
z_{\bar{w}}=\mu z_{w},\label{eq:BeltramiCoefficient}
\end{equation}
where $\mu$ is called the \textit{Beltrami coefficient}. This is
a very useful quantity as it measures how far a quasiconformal map
is from the $\mu=0$ conformal case: Its modulus determines the ratio
of the principal stretches, $\mathcal{K}=(1+\left|\mu\right|)/(1-\left|\mu\right|)$,
while its argument describes the orientation of the stretching direction
in the reference plane. The local area dilation is,
\begin{equation}
\mathcal{J}=\left|z_{w}\right|^{2}-\left|z_{\bar{w}}\right|^{2}=\left|z_{w}\right|^{2}\left(1-\left|\mu\right|^{2}\right),
\end{equation}
leading to a natural condition that the map is quasiconformal if $\left|\mu\right|<1$
with $\mu=0$ recovering the conformal case. Hence, a quasiconformal
map encodes local scaling, rotation and stretching at once. 

For the stretch example, $z_{w}=\frac{1}{2}(\alpha+1)$ and $z_{\bar{w}}=\frac{1}{2}(\alpha-1)$,
so that, 
\begin{equation}
\mu=\frac{z_{\bar{w}}}{z_{w}}=\frac{\alpha-1}{\alpha+1},\label{eq:stretchBeltrami}
\end{equation}
and the area dilation is simply $\mathcal{J}=\alpha$. It is instructive
to examine some limiting cases: as $\alpha\to0$, $\mu\to-1$, while
as $\alpha\to\infty$, $\mu\to+1$. Hence the map is quasiconformal
for any value of $\alpha>0$ .

\section{Anisotropic elastic geometry}\label{sec:Anisotropic-elastic-geometry}

Returning to the Frank energy, Eq. (\ref{eq:Frank2D}) and our parameterization
of the director $\mathbf{n}=\left(\cos\theta,\sin\theta\right)$,
we evaluate the divergence $\nabla\cdot\mathbf{n}=\mathbf{n}\times\nabla\theta$
and curl $\nabla\times\mathbf{n}=\mathbf{n}\cdot\nabla\theta$ and
write the energy as a quadratic form,
\begin{equation}
F=\frac{1}{2}\int_{\Omega}d^{2}x\left[\nabla\theta^{T}\cdot A(\theta)\cdot\nabla\theta\right],\label{eq:quadraticform}
\end{equation}
where $A(\theta)$ is the elastic anisotropy tensor, 
\begin{equation}
A(\theta)=K_{1}I_{2}+\left(K_{3}-K_{1}\right)\mathbf{n}\otimes\mathbf{n},\label{eq:lcelasticmetric}
\end{equation}
with $I_{2}$ as the $2\times2$ identity matrix and $\otimes$ denotes
the usual outer product. Note that $A(\theta)$ is spatially varying
because it depends on the local orientation of the director. It is
real and symmetric with eigenvalues $K_{1}$ and $K_{3}$; if $K_{1}>0$
and $K_{3}>0$ it is also positive definite. It may be decomposed
into $A=R(\theta)\text{diag}(K_{3},K_{1})R^{T}(\theta)$, where $R(\theta)$
is a 2D rotation matrix, providing a geometric interpretation: it
encodes a \textit{local} instruction to stretch by a factor determined
by the elastic constants in a rotated frame specified by the director
angle. 

It is natural to ask if the anisotropy can be removed by a suitable
change of coordinates, and the geometric picture immediately suggests
that a quasiconformal map might be able to do so. We therefore consider
a map $z(w)$ with Jacobian $J=\frac{\partial(x,y)}{\partial(u,v)}$
and treat Eq. (\ref{eq:quadraticform}) as written in the physical
plane. Letting $\theta(z)=\phi(w)$ and defining $A(w)=A(z(w))$,
we may transform Eq. (\ref{eq:quadraticform}) to the reference $w$-plane,
\begin{equation}
F=\frac{1}{2}\int_{\Omega_{w}}d^{2}w\det J\left[\nabla_{w}\phi^{T}J^{-1}\cdot A(w)\cdot J^{-T}\nabla_{w}\phi\right],
\end{equation}
where $\nabla_{w}$ is the gradient in the reference plane. If the
map can satisfy,
\begin{equation}
\det J\left(J^{-1}\cdot A(w)\cdot J^{-T}\right)=\lambda I_{2},\label{eq:mapcondition}
\end{equation}
where $\lambda$ is a scalar, then the energy in the $w$-plane becomes
isotropic,
\begin{equation}
F=\frac{1}{2}\int_{\Omega_{w}}d^{2}w\lambda\left|\nabla_{w}\phi\right|^{2}.\label{eq:isotropicwform}
\end{equation}

The condition Eq. (\ref{eq:mapcondition}) necessitates the full generality
of a quasiconformal map as it includes local scaling, rotation and
stretching. Rearranging Eq. (\ref{eq:mapcondition}),
\begin{equation}
A=\frac{\lambda}{\det J}JJ^{T},\label{eq:AfromJ}
\end{equation}
we relate the elastic anisotropy tensor directly to the quasiconformal
deformation induced by the map. Hence, anisotropic elasticity may
be interpreted geometrically as a rotating field of local quasiconformal
distortions. We may take the determinant of Eq. (\ref{eq:AfromJ})
to arrive at the relation $\lambda=\sqrt{\det A}$ and use this to
define an elastic metric,
\begin{equation}
g=\sqrt{\det A}A^{-1},
\end{equation}
encoding the quasiconformal geometry associated with the anisotropic
energy. 

Unlike the isotropic case, where the Euclidean metric is fixed, anisotropic
elasticity defines a spatially varying metric whose principal axes
rotate with the nematic director. The geometry therefore depends on
the director field itself and becomes part of the variational problem.
Although a suitable quasiconformal transformation isotropizes the
elastic quadratic form for a given director configuration, it does
not generally reduce the full liquid-crystal variational problem to
the one-constant form Eq. (\ref{eq:isotropicwform}). We return to
this distinction in subsections \ref{subsec:Quasiconformal-maps-for}
and \ref{subsec:Hybrid-aligned-liquid-crystal} below.

\subsection{Quasiconformal maps for LC problems}\label{subsec:Quasiconformal-maps-for}

Returning to the nematic elastic anisotropy tensor, Eq. (\ref{eq:lcelasticmetric}),
we first introduce its square root,
\begin{equation}
U(\theta)=A^{1/2}(\theta)=\sqrt{K_{1}}I_{2}+\left(\sqrt{K_{3}}-\sqrt{K_{1}}\right)\mathbf{n}\otimes\mathbf{n},\label{eq:lcjacobian}
\end{equation}
which can be decomposed $U(\theta)=R(\theta)\Lambda R^{T}(\theta)$,
where $\Lambda=\text{diag}(\sqrt{K_{3}},\sqrt{K_{1}})$ is the stretch
in the diagonal frame and $\det U=\sqrt{K_{1}K_{3}}$. The matrix
$U$ satisfies Eq. (\ref{eq:mapcondition}) with $\lambda=\sqrt{K_{1}K_{3}}$,
\begin{equation}
\left(\det U\right)U^{-1}AU^{-T}=\sqrt{K_{1}K_{3}}I_{2},
\end{equation}
and hence provides the exact local anisotropic stretch required to
render the elastic quadratic form isotropic. A very attractive property
of the liquid crystal problem is that the principal stretches $\sqrt{K_{1}}$
and $\sqrt{K_{3}}$ are spatially constant; only the principal directions
rotate with the director. Hence, the magnitude of elastic anisotropy
is encoded homogeneously, while its orientation varies spatially.

To construct an isotropizing map, it is necessary to compose $U$
with an overall isotropic dilation $s(w)$ and a local rotation parametrized
by $\gamma(w)$,
\begin{equation}
J=s(w)U(\phi)R(\gamma(w)),\label{eq:lcnewjacobian}
\end{equation}
so that $JJ^{T}=s^{2}U^{2}=s^{2}A(\phi)$ and $\det J=s^{2}(w)\det U$.
For such a map to exist, the mixed derivatives must commute,
\begin{equation}
x_{uv}=x_{vu},\ y_{uv}=y_{vu},\label{eq:compatibility}
\end{equation}
a condition we henceforth refer to as compatibility, leading to a
system of equations for $s(w)$ and $\gamma(w)$.

In complex notation, scaling and rotation are written $s(w)\left[\zeta\right]=s\zeta$
and $R(\gamma)\left[\zeta\right]=e^{i\gamma}\zeta$ respectively,
where $\zeta=\zeta^{x}+i\zeta^{y}$ is an arbitrary complex number
representing a planar vector $(\zeta^{x},\zeta^{y})$. The stretch
is $\Lambda\left[\zeta\right]=\sqrt{K_{3}}\zeta^{x}+i\sqrt{K_{1}}\zeta^{y}=k\zeta+\Delta k\bar{\zeta}$,
paralleling the earlier example Eq. (\ref{eq:stretchcomplex}). Hence,
\begin{align}
U(\phi)\left[\zeta\right] & =R(\phi)\circ\Lambda\circ R^{-1}(\phi)\left[\zeta\right]\nonumber \\
 & =k\zeta+\Delta ke^{i2\phi}\bar{\zeta}.\label{eq:finalUansatz}
\end{align}
Composing these, we construct the candidate Jacobian Eq. (\ref{eq:lcnewjacobian}),
\begin{align}
J\left[\zeta\right] & =s\circ U(\phi)\circ R(\gamma)\left[\zeta\right]\nonumber \\
 & =ske^{i\gamma}\zeta+s\Delta ke^{i2\phi-i\gamma}\overline{\zeta}.\label{eq:CandidateJ}
\end{align}
Comparing Eq. (\ref{eq:lcnewjacobian}) with the standard differential
of a quasiconformal map,
\begin{equation}
J\left[\zeta\right]=z_{w}\zeta+z_{\bar{w}}\bar{\zeta},
\end{equation}
we read off $z_{w}$ and $z_{\bar{w}}$,
\begin{equation}
z_{w}=ske^{i\gamma},\ z_{\bar{w}}=s\Delta ke^{i2\phi-i\gamma},\label{eq:zwzwbardefn}
\end{equation}
and note the relation,
\begin{equation}
z_{\bar{w}}=\xi e^{i2\phi}\overline{z_{w}},\label{eq:zwbartobarzw}
\end{equation}
where,
\begin{equation}
\xi=\frac{\Delta k}{k}=\frac{\sqrt{K_{3}}-\sqrt{K_{1}}}{\sqrt{K_{3}}+\sqrt{K_{1}}}.\label{eq:xidefinition}
\end{equation}

The compatibility condition Eq. (\ref{eq:compatibility}) in complex
notation is,
\begin{equation}
\partial_{\bar{w}}z_{w}=\partial_{w}z_{\bar{w}}.\label{eq:ComplexCompatibility-1}
\end{equation}
and inserting Eq. (\ref{eq:zwzwbardefn}) we obtain,
\begin{align}
k\partial_{\bar{w}}\left(se^{i\gamma}\right) & =\Delta k\partial_{w}\left(se^{i2\phi-i\gamma}\right),\\
k\left(e^{i\gamma}s_{\bar{w}}+ise^{i\gamma}\gamma_{\bar{w}}\right) & =\Delta ke^{i2\phi-i\gamma}\left[s_{w}+is\left(2\phi_{w}-\gamma_{w}\right)\right]
\end{align}
Dividing through by $se^{i\gamma}$ yields, 
\begin{equation}
\left(\log s\right)_{\bar{w}}+i\gamma_{\bar{w}}=\frac{\Delta k}{k}e^{i2\left(\phi-\gamma\right)}\left[\left(\log s\right)_{w}+i\left(2\phi_{w}-\gamma_{w}\right)\right],\label{eq:sgammacompatibility}
\end{equation}
which is a complex equation, for the two real unknown fields $s(w)$
and $\gamma(w)$. If these can be solved, a quasiconformal map with
the desired Jacobian can be constructed with a corresponding Beltrami
coefficient,
\begin{equation}
\mu=\xi e^{i2\left(\phi-\gamma\right)}.\label{eq:BeltramiLCNew}
\end{equation}
Note that the signed prefactor $\xi$ directly encodes the elastic
constants, and acts as a natural measure of elastic anisotropy in
the quasiconformal picture. Meanwhile, the phase of the Beltrami coefficient
Eq. (\ref{eq:BeltramiLCNew}) encodes the director angle measured
relative to the locally rotated quasiconformal frame. Writing the
director in complex form as $n=\cos\theta+i\sin\theta=e^{i\theta}$,
we see $\mu=\xi n^{2}e^{-2i\gamma}$. The square is consistent with
the symmetry of the nematic phase $n\to-n$, so that $n$ and $-n$
both yield the same stretching axis. 

We conclude by noting an additional consequence of Eq. (\ref{eq:CandidateJ}):
while the stretch operator $U$ locally isotropizes the elastic quadratic
form, the rotation field $\gamma(w)$ also defines a natural connection
one-form $\omega=d\gamma$ associated with the locally rotating quasiconformal
frame. Writing the director angle relative to this frame as $\vartheta=\phi-\gamma$,
we obtain $d\phi=d\vartheta+\omega$, and the transformed energy takes
the frame-covariant form,
\begin{equation}
F=\frac{\sqrt{K_{1}K_{3}}}{2}\int_{\Omega_{w}}\left|d\vartheta+\omega\right|^{2}d^{2}w.\label{eq:EnergyFrameCovariant}
\end{equation}
Thus while the explicit directional elasticity is removed, the spatial
rotation required to assemble the local stretches into a compatible
map survives through the frame connection. We provide explicit examples
of such connections below. We also note recent work by Pollard and
Morris\citep{PollardMorris2026} who developed a geometric field theory
of two-dimensional nematic defects, and showed that elastic anisotropy
cannot be expressed solely in terms of the director connection. Here,
the quasiconformal transformation and its associated frame connection
$\omega$ allows the anisotropic energy to be expressed in the frame-covariant
form Eq. (\ref{eq:EnergyFrameCovariant}). 

\subsection{Hybrid-aligned liquid crystal film}\label{subsec:Hybrid-aligned-liquid-crystal}

\begin{figure*}
\includegraphics[width=1\textwidth]{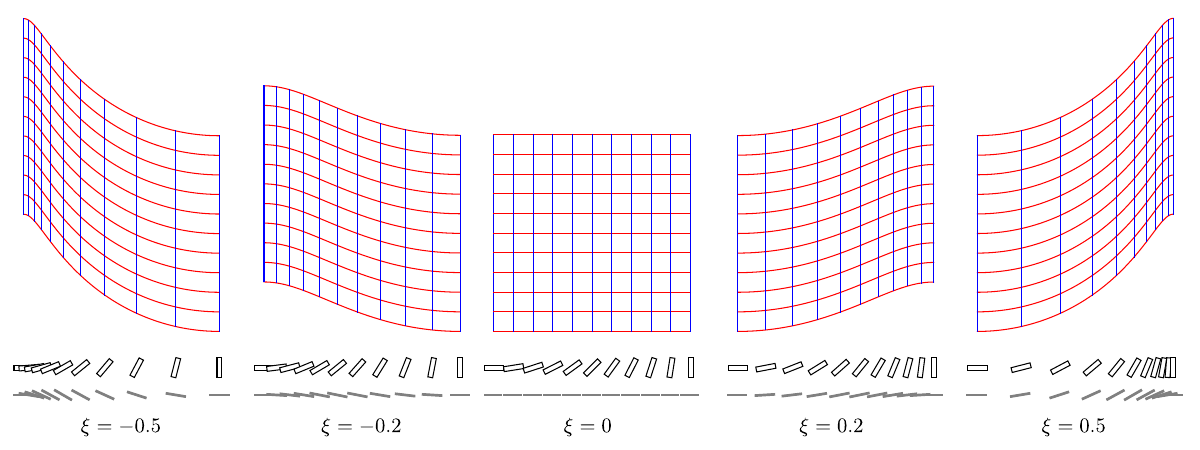}

\caption{\textbf{Quasiconformal isotropization of the hybrid-aligned nematic
texture.} Images of a rectangular grid in the auxiliary $w$-plane
under the isotropizing map are shown for several values of the elastic
anisotropy parameter $\xi=\frac{\sqrt{K_{3}}-\sqrt{K_{1}}}{\sqrt{K_{3}}+\sqrt{K_{1}}}$;
red curves are the images of lines of constant $v$, while blue curves
are the images of lines of constant $u$. Also shown are director
profiles (black rectangles) and the local polar rotation $\gamma=\arg z_{w}$
induced by the map (gray rectangles). Note that map reduces to the
identity operation in the isotropic case $\xi=0$, while changing
the sign of $\xi$ reverses the direction of the induced shear.}\label{fig:HANmaps}
\end{figure*}

To explore the ramifications of a $\theta$-dependent elastic metric,
we first consider a very simple one-dimensional problem, the structure
of a hybrid-aligned nematic (HAN) film. The domain is $x\in[-L/2,L/2]$
with boundary conditions $\theta(-L/2)=0$ and $\theta(L/2)=\pi/2$.
The Frank energy per unit length Eq. (\ref{eq:Frank2D}) becomes,
\begin{equation}
F=\frac{1}{2}\int_{-L/2}^{L/2}P(\theta)\theta'{}^{2}dx,\label{eq:Frank1D}
\end{equation}
where $P(\theta)=A_{xx}(\theta)=K_{1}\sin^{2}\theta+K_{3}\cos^{2}\theta$.
When $K=K_{1}=K_{3}$, the energy reduces to,
\begin{equation}
F=\frac{K}{2}\int_{-L/2}^{L/2}\theta_{x}{}^{2}dx.\label{eq:Frank1D-1}
\end{equation}
and the minimizer is a linear function,
\begin{equation}
\theta(x)=\frac{\pi}{2}\frac{x+L/2}{L}.\label{eq:isotropicminimizer}
\end{equation}
For anisotropic elasticity, an implicit solution exists in terms of
elliptic integrals,
\begin{equation}
E(\pi/2-\theta\mid m)=E(m)\left(\frac{1}{2}-\frac{x}{L}\right),
\end{equation}
where $E(\varphi\mid m)=\int_{0}^{\varphi}\sqrt{1-m\sin^{2}t}dt$
is the incomplete elliptic function of the second kind where here
$m=1-\frac{K_{3}}{K_{1}}$ is the elliptic parameter.

We now show a transformation that reduces Eq. (\ref{eq:Frank1D})
to a harmonic form. Defining a new angular variable as follows,
\begin{equation}
\Psi(\theta)=\int_{0}^{\theta}\sqrt{P(t)}dt,\label{eq:NewPsi1D}
\end{equation}
the energy Eq. (\ref{eq:Frank1D}) can be rewritten
\begin{equation}
F=\frac{1}{2}\int_{-L/2}^{L/2}\left(\frac{d\Psi}{dx}\right)^{2}dx,\label{eq:FPsiMinimizer}
\end{equation}
equivalent to the one-constant form Eq. (\ref{eq:Frank1D-1}) in the
new variable $\Psi$. Note that the transformation Eq. (\ref{eq:NewPsi1D})
deforms the angle $\theta$, not the spatial coordinate. However,
the minimizer of Eq. (\ref{eq:FPsiMinimizer}) must be a linear function.
Applying the boundary conditions on $\theta$ and rearranging, we
obtain an expression for $x$,
\begin{equation}
x(\theta)=-\frac{L}{2}+L\frac{\Psi(\theta)}{\Psi(\pi/2)}.\label{eq:HANxfnu}
\end{equation}
We may also introduce an auxiliary coordinate $\tilde{u}$, in which
the angular solution is a linear function, $\Phi(\tilde{u})=\frac{\pi}{2}(\tilde{u}+L/2)/L$,
and identify this as pullback of $\theta$ onto the $\tilde{u}$ coordinate,
$\theta(x(\tilde{u}))=\Phi(\tilde{u}).$ We can use this to construct
$x(\tilde{u})$,
\begin{equation}
x(\tilde{u})=-\frac{L}{2}+L\frac{\Psi(\Phi(\tilde{u}))}{\Psi(\pi/2)},\label{eq:HANtransform}
\end{equation}
and, by integrating Eq. (\ref{eq:NewPsi1D}) obtain an explicit expression,
\begin{equation}
x(\tilde{u})=\frac{L}{2}-L\frac{E(\pi/2-\Phi(\tilde{u})\mid m)}{E(m)}.\label{eq:map1D}
\end{equation}
We have therefore obtained an explicit one-dimensional map $x(\tilde{u})$
whose pullback of the anisotropic solution $\theta(x)$ is the trivial
linear field $\Phi(\tilde{u})$.

We now seek a different auxiliary coordinate transformation $z(w)$
that isotropizes the full two-dimensional elastic tensor. Exploiting
the property that the HAN solution is one-dimensional, we seek a map
where $z_{w}$ and $z_{\bar{w}}$ only depend on $u$. Setting $z_{w}=p(u)$,
compatibility requires,
\begin{equation}
\partial_{\bar{w}}z_{w}-\partial_{w}z_{\bar{w}}=\frac{1}{2}p_{u}-\frac{1}{2}\left(z_{\bar{w}}\right)_{u}=0,
\end{equation}
which can be integrated to yield $p-z_{\bar{w}}=C$. Choosing $C=1$
as an overall normalization implies $z_{v}=i$, and using the relation
Eq. (\ref{eq:zwbartobarzw}) we solve for $p$,
\begin{equation}
p=z_{w}=\frac{1+\xi e^{2i\phi}}{1-\xi^{2}},\ z_{\bar{w}}=\frac{\xi^{2}+\xi e^{2i\phi}}{1-\xi^{2}},
\end{equation}
and combine these to obtain,
\begin{equation}
z_{u}=\frac{P(\phi)+iQ(\phi)}{\lambda},\label{eq:HANzu}
\end{equation}
where $P(\phi)=K_{1}\sin^{2}\phi+K_{3}\cos^{2}\phi$, $Q(\phi)=\left(K_{3}-K_{1}\right)\sin\phi\cos\phi$
and $\lambda=\sqrt{K_{1}K_{3}}$ as before. 

The map can then be written,
\begin{equation}
z(w)=\int^{u}\frac{P(\phi(t))+iQ(\phi(t))}{\lambda}dt+iv.\label{eq:FinalHANmap}
\end{equation}
As discussed above, as well as the quasiconformal distortion, the
map Eq. (\ref{eq:FinalHANmap}) induces a local conformal dilation,
\begin{align}
\rho=\left|z_{w}\right|=sk & =k\sqrt{P(\phi)}/\lambda\label{eq:HANdilation}
\end{align}
and a local rotation,
\begin{align}
\gamma=\arg z_{w} & =\arg\left(P(\phi)+\lambda+iQ(\phi)\right),\label{eq:HANrotation}
\end{align}
that together satisfy the compatibility relation Eq. \ref{eq:sgammacompatibility}. 

We also can use the map Eq. (\ref{eq:FinalHANmap}) to illuminate
the frame connection introduced earlier. Differentiating Eq. (\ref{eq:HANrotation})
using $d\arg Z=\Im\left(dZ/Z\right)$ we obtain,
\begin{align}
\omega=d\gamma & =\frac{\left(P+\lambda\right)dQ-QdP}{\left(P+\lambda\right)^{2}+Q^{2}}\nonumber \\
 & =\left(1-\frac{\lambda}{P(\phi)}\right)d\phi.\label{eq:HANspin}
\end{align}
Since $\gamma$ is smooth and single-valued in this geometry, the
connection $\omega=d\gamma$ is pure-gauge and curvature free: it
simply records the rotation of the frame. 

To evaluate the map numerically, we must first determine the pulled-back
director profile $\phi(u)=\theta(x(u))$. Differentiating the HAN
equilibrium solution Eq. (\ref{eq:HANxfnu}) obtained above and combining
it with the real part of Eq. (\ref{eq:HANzu}), $dx/du=P(\phi)/\lambda$,
i.e. the longitudinal stretch, we obtain,

\begin{equation}
\frac{d\phi}{du}=\frac{d\theta}{dx}\frac{dx}{du}=\frac{\Psi(\pi/2)}{\lambda L}\sqrt{P(\phi)},
\end{equation}
which can be integrated and inverted to obtain $\phi(u)$. Since $Q(\phi)=-\frac{1}{2}P'(\phi)$,
the integral in (\ref{eq:FinalHANmap}) can be evaluated, 
\begin{align}
z(u+iv)= & -\frac{L}{2}+\frac{L\Psi(\phi(u))}{\Psi(\pi/2)}\label{eq:Znumerical}\\
 & +i\left[v+\frac{L}{\Psi(\pi/2)}\left(\sqrt{K_{3}}-\sqrt{P(\phi(u))}\right)\right],\nonumber 
\end{align}
where constants of integration were chosen so that the left boundary
maps to $x=-L/2$.

Using the formula Eq. (\ref{eq:Znumerical}), the image of the auxiliary
$w$-plane under the map Eq. (\ref{eq:FinalHANmap}) is plotted for
several values of $\xi$, the signed anisotropy parameter, in Fig.
\ref{fig:HANmaps}, together with the associated director profile
and the rotation induced by the quasiconformal frame $\gamma$. We
stress that the auxiliary coordinate used here isotropizes the complete
two-dimensional tensor; it does not in general transform the anisotropic
solution into a linear director profile in the auxiliary coordinate
as Eq. (\ref{eq:map1D}) does.

\section{Anisotropic defect structure}\label{sec:Anisotropic-defect-structure}

We now seek to discover the quasiconformal analog of the conformal
defect construction introduced in section \ref{sec:ConformalDefects}.
The familiar radial or azimuthal $+1$ defects are respectively pure
splay or bend. Hence, anisotropic elasticity does not change their
structure, although it does change the material's relative energetic
preference for each type. We therefore examine the structure of defects
with different charges and begin by reviewing the classical Dzyaloshinskii
solution.

Remarkably, the solution is self-similar: written using polar coordinates
$x+iy=re^{i\varphi}$, there is no $r$ dependence, and instead the
dependence is solely on $\varphi$. In the below we have changed the
role of $\theta$ and $\phi$ relative to that in Ref. \citep{hudson1989frank};
here $\theta$ is the director angle and $\varphi$ is the spatial
coordinate in polar coordinates. The symbol $\phi$ is reserved for
the director angle defined on an auxiliary domain.

Defining $\epsilon$ as an anisotropy parameter, 
\begin{equation}
\epsilon=\frac{K_{1}-K_{3}}{K_{1}+K_{3}},
\end{equation}
noting that $-1<\epsilon<1$, then the solution is given by
\begin{equation}
\varphi=p\int_{0}^{\theta-\varphi}\left[\frac{1+\epsilon\cos2t}{1+p^{2}\epsilon\cos2t}\right]^{\frac{1}{2}}dt,\label{eq:Dzyaloshinskii}
\end{equation}
where $p$ is a constant determined by imposing the prescribed defect
winding, $\Delta\theta=2\pi s,$ upon one circuit of the defect so
that,
\begin{equation}
(s-1)p\int_{0}^{\pi}\left[\frac{1+\epsilon\cos2t}{1+p^{2}\epsilon\cos2t}\right]^{\frac{1}{2}}dt=\pi,\label{eq:closure}
\end{equation}
where $s$ is the defect strength; here $s=+1/2$. To obtain a texture
we first find a value of $p$ that solves the condition Eq. (\ref{eq:closure})
and then successively insert director angles of interest into Eq.
(\ref{eq:Dzyaloshinskii}) and solve for the angular coordinate $\varphi$
at which they occur. 

\subsection{A geometrically motivated auxiliary domain}

We now show a surprising connection between Eq. (\ref{eq:Dzyaloshinskii})
and our one dimensional example in Section \ref{subsec:Hybrid-aligned-liquid-crystal}.
Defining,
\begin{equation}
\Xi(\beta)=\int_{0}^{\beta}\left[\frac{1+\epsilon\cos2t}{1+p^{2}\epsilon\cos2t}\right]^{\frac{1}{2}}dt,\label{eq:DzyaloshinskiiAngle}
\end{equation}
so that Eq. (\ref{eq:Dzyaloshinskii}) is just,
\begin{equation}
\varphi=p\Xi(\theta-\varphi),\label{eq:DzyaloshinskiiEqNew}
\end{equation}
we notice that Eq. (\ref{eq:Dzyaloshinskii}) is a reparametrization
of the angular coordinate, paralleling the reparametrization Eq. (\ref{eq:NewPsi1D})
in the HAN example. Similarly, Eq. (\ref{eq:DzyaloshinskiiEqNew})
parallels the earlier map $x(u)$ in Eq. (\ref{eq:HANtransform}).
This parallel immediately provides a new geometric object embedded
in the Dzyaloshinskii solution. 

In Eq. (\ref{eq:HANtransform}) the quantity $\Psi(\Phi(u))$ may
be interpreted as an intermediate auxiliary coordinate in building
the map $u\to\Phi\to\Psi\to x$. Its range is naturally $0\le\Psi(\Phi(u))\le\Psi(\pi/2)$
and hence $\Psi(\pi/2)$ may be interpreted as the natural total arclength
of the intermediate auxiliary domain, which depends on the elastic
anisotropy. The term,
\begin{equation}
L\frac{\Psi(\Phi(u))}{\Psi(\pi/2)},\label{eq:HANrescaling}
\end{equation}
is a uniform global rescaling chosen to map the image of $\Psi$ onto
the physical domain with length $L$. Since the domain is a line segment,
there is no impediment to this rescaling.

Rewriting the condition Eq. (\ref{eq:closure}) using the definition
Eq. (\ref{eq:DzyaloshinskiiAngle}) and setting the defect strength
$s=+1/2$,
\begin{equation}
-\frac{1}{2}p\Xi(\pi)=\pi,\label{eq:ClosureNew}
\end{equation}
then dividing Eq. (\ref{eq:DzyaloshinskiiEqNew}) by (\ref{eq:ClosureNew})
and rearranging we obtain,
\begin{equation}
\varphi=-2\pi\frac{\Xi(\theta-\varphi)}{\Xi(\pi)}=p\Xi(\theta-\varphi).
\end{equation}
The parallel with Eq. (\ref{eq:HANrescaling}) is now clear: $2\pi/\Xi(\pi)=-p$
is a global angular rescaling that maps the image of $\Xi$ onto the
angular interval $2\pi$. In contrast to the open linear case, however,
rescaling a periodic interval leads to nontrivial phenomena: Up to
the choice of orientation, the domain parametrized by $\Xi$ is naturally
a wedge of angle $\Omega=\Xi(\pi)$ but with radial lines at $\Xi=0$
and $\Xi=\Xi(\pi)$ identified; geometrically such a construction
is a cone. The quantity $p$ in Dzyaloshinskii's solution acquires
a new geometric significance in defining the cone angle $\Omega$
of an intermediate auxiliary domain. 

Nematic and more general $p$-atic order on conical substrates has
previously been studied by Nelson and collaborators, who showed how
the apex curvature and deficit angle act as a geometric charge and
influence defect structure, absorption and emission\citep{ZhangNelson2022,VafaZhangNelson2022}.
While this connection is pleasing, we emphasize that the cone here
is not a physical substrate, but an intrinsic auxiliary geometry generated
by elastic anisotropy and the global closure condition.

In the isotropic limit, when $\epsilon\to0$, the definition of $\Xi$
gives $\Xi(\beta)\to\beta$ and hence $\Omega=\Xi(\pi)\to\pi$ and
the closure condition Eq. (\ref{eq:closure}) gives $p=-2$. The auxiliary
domain in this case is simply the half plane $\mathbb{H}$ with the
positive and negative real axes identified, precisely the reference
domain used in the conformal construction of the $+1/2$ defect texture
in Section \ref{sec:ConformalDefects}. 

\subsection{Quasiconformal construction of anisotropic defects}

\begin{figure*}
\includegraphics[width=0.9\textwidth]{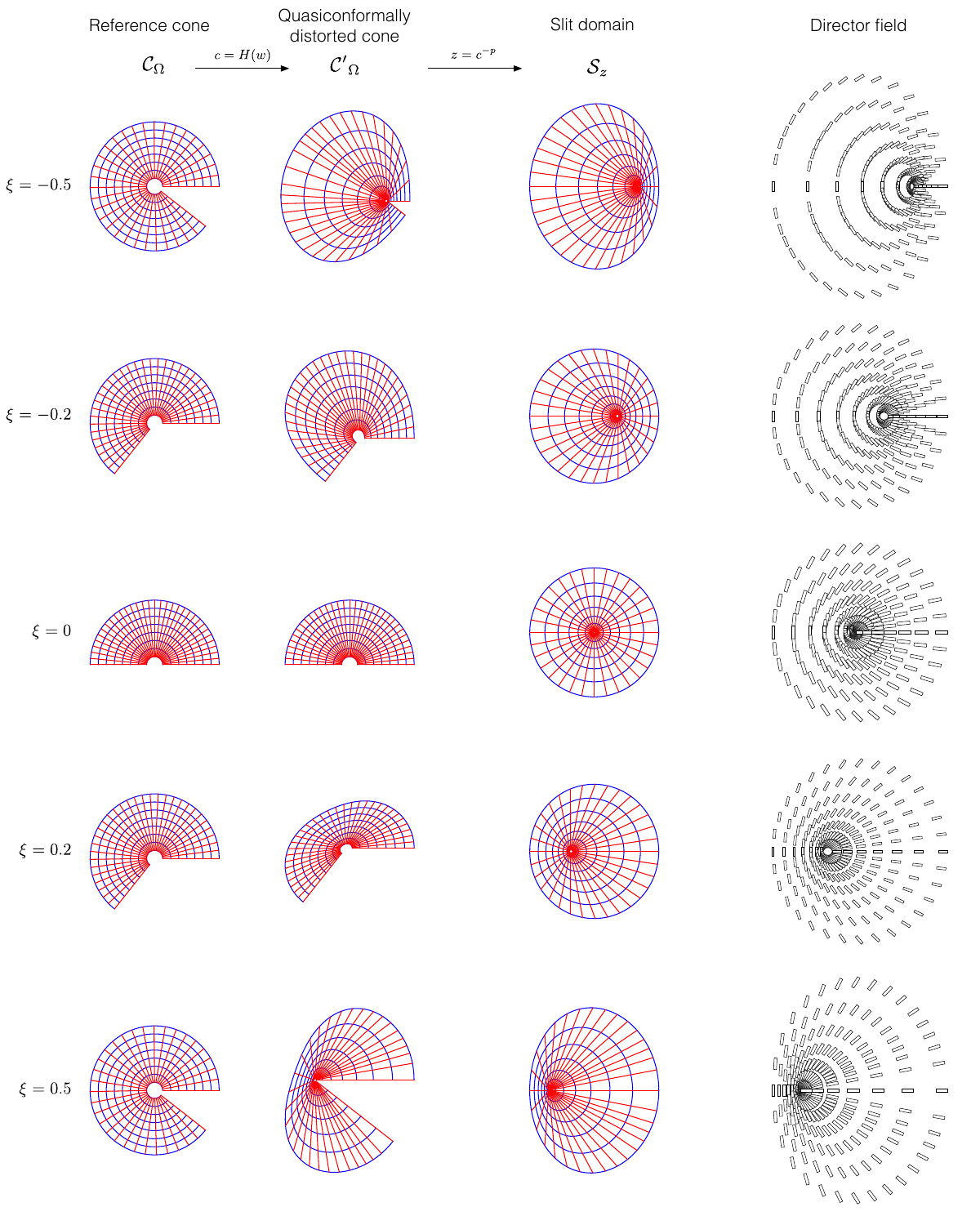}

\caption{\textbf{Quasiconformal construction of the anisotropic $+\frac{1}{2}$
defect.} The defect is constructed as the composition of a quasiconformal
map between two cones $\mathcal{C}_{\Omega}$ (left) and $\mathcal{C}'_{\Omega}$
(center) of identical opening angle $\Omega$, followed by a conformal
power map onto the physical slit domain $S_{z}$ (right). The image
of a polar grid in the reference domain is shown projected into each
domain as a function of the anisotropy parameter $\xi$. In the isotropic
limit $\xi\to0$, the quasiconformal map reduces to the identity and
the construction becomes the $z=w^{2}$ conformal construction of
the defect presented in Section \ref{sec:ConformalDefects}.}\label{fig:DefectMaps}
\end{figure*}

Equipped with the insights of the previous section, we will now show
that the Dzyaloshinskii solution for the $+\frac{1}{2}$ defect can
be constructed using a sequence of maps between domains,
\begin{equation}
\mathcal{C}_{\Omega}\xrightarrow{c=H(w)}\mathcal{C}'_{\Omega}\xrightarrow{z=c^{-p}}\mathcal{S}_{z}.\label{eq:TwoStepMap}
\end{equation}
Here $\mathcal{C}_{\Omega}$ is the reference cone of angle $\Omega$
identified in the previous subsection parametrized by a coordinate
$w$; this is mapped through a quasiconformal transformation $H(w)$
onto a second cone $\mathcal{C}'_{\Omega}$ parametrized by $c$,
with the same opening angle $\Omega$ and seam identification but
a distorted coordinate geometry. The cone $\mathcal{C}'_{\Omega}$
is then wrapped into the slit domain $\mathcal{S}_{z}$ with coordinate
$z$ using a conformal power map $z=c^{-p}$. The quantity $p$ is
that defined by the closure condition in Eq. (\ref{eq:ClosureNew}).
The two bounding rays of $\mathcal{C}'_{\Omega}$ represent the same
geometric seam, but remain on distinct sides of the branch cut, just
as the two sides of the slit remain distinct in $\mathcal{S}_{z}$.

The composition of these maps separates the quasiconformal distortion,
encoded in $H$, associated with elastic anisotropy from the conformal
transformation that produces the physical defect geometry and contributes
only local scaling and rotation. In the isotropic limit, $C_{\Omega}$
is $\mathbb{H}$ and the map $H(w)$ becomes the identity map; the
construction Eq. (\ref{eq:TwoStepMap}) reduces to $\mathbb{H}\xrightarrow{z=w^{2}}\mathcal{S}_{z}$
as in Section \ref{sec:ConformalDefects}. We therefore can interpret
Eq. (\ref{eq:TwoStepMap}) as the anisotropic generalization of the
conformal construction.

It remains to construct the map $c=H(w)$. To do so, we write $w$,
$c$ and $z$ in polar form so that,
\begin{equation}
w=Re^{i\eta}\in\mathcal{C}_{\Omega},\ c=\mathcal{R}e^{i\psi}\in\mathcal{C}'_{\Omega},\ z=c^{q}=re^{i\varphi}\in\mathcal{S}_{z},
\end{equation}
where we defined $q=-p>0$ and $q\Omega=2\pi$. We first transport
the director back from the physical domain to the $c$-plane. Since
the power map rotates an arbitrary vector by,
\begin{align}
\arg z_{c} & =\arg\left(qc^{q-1}\right)=(q-1)\psi\label{eq:RotationFromPowerMap}
\end{align}
the director angle pulled back onto the $c$-cone is $\Theta=\theta-(q-1)\psi,$
and since the polar angle is transported,
\begin{equation}
\varphi=q\psi,\label{eq:pullbackOfAngularCoordinate}
\end{equation}
we obtain,
\begin{equation}
\Theta-\psi=\theta-\varphi=\beta,
\end{equation}
corresponding to the relative angle the director makes with the radial
direction on the cone $\Theta-\psi$ and in the physical plane $\theta-\varphi$;
this is preserved at corresponding points in both domains since the
power map is conformal. 

The compatibility condition for the $c=H(w)$ map has the same form
as Eq. (\ref{eq:zwbartobarzw}), but must be written using the director
angle $\Theta$ in the target $c$ plane,
\begin{equation}
c_{\bar{w}}=\xi e^{i2\Theta}\overline{c_{w}}=\xi e^{i2\left(\beta+\psi\right)}\overline{c_{w}},\label{eq:CompatabilityForC}
\end{equation}
where, $\xi$ is the anisotropy parameter defined in Eq. (\ref{eq:xidefinition}).
To solve this, we posit a self-similar ansatz,
\begin{equation}
c=H(w)=R^{\kappa}e^{h(\eta)+i\psi(\eta)},
\end{equation}
so that the map determines an angular parametrization $\eta\to\psi(\eta)$,
compensated for by a radial distortion; this form is motivated by
the radial independence of the Dzyaloshinskii solution. Using the
polar form of the derivative operator, 
\begin{equation}
\partial_{w}=\frac{e^{-i\eta}}{2}\left(\partial_{R}-\frac{i}{R}\partial_{\eta}\right),
\end{equation}
we evaluate the derivatives of $c$,
\begin{align}
c_{w} & =\frac{ce^{-i\eta}}{2R}\left(\kappa+\psi_{\eta}-ih_{\eta}\right),\\
c_{\bar{w}} & =\frac{ce^{i\eta}}{2R}\left(\kappa-\psi_{\eta}+ih_{\eta}\right).
\end{align}
Substituting these into the compatibility condition Eq. (\ref{eq:CompatabilityForC})
and using $c=\mathcal{R}e^{i\psi}$ and $\bar{c}=\mathcal{R}e^{-i\psi}$
to obtain the relation $c=e^{2i\psi}\bar{c}$, we obtain,
\begin{equation}
\left(\kappa-\psi_{\eta}+ih_{\eta}\right)=\xi e^{i2\beta}\left(\kappa+\psi_{\eta}+ih_{\eta}\right).\label{eq:CompatabilityWithAnsatz}
\end{equation}
Solving the real and imaginary parts of Eq. (\ref{eq:CompatabilityWithAnsatz})
and using the following relation between the Dzyaloshinskii anisotropy
parameter and the Beltrami coefficient,
\begin{equation}
\epsilon=-\frac{2\xi}{1+\xi^{2}},
\end{equation}
we arrive at two ordinary differential equations,
\begin{align}
\psi_{\eta} & =\frac{\kappa}{\sqrt{1-\epsilon^{2}}}\left(1+\epsilon\cos2\beta\right),\label{eq:CompatibilityPsi}\\
h_{\eta} & =-\frac{\kappa}{\sqrt{1-\epsilon^{2}}}\epsilon\sin2\beta.\label{eq:CompatibilitySystem}
\end{align}
Equations (\ref{eq:CompatibilityPsi}) and (\ref{eq:CompatibilitySystem})
reduce the compatibility problem to two coupled first-order equations.
In the next section we show that the Dzyaloshinskii solution supplies
a globally compatible solution.

\subsection{Explicit quasiconformal map}

\begin{figure*}
\begin{centering}
\includegraphics[width=1\textwidth]{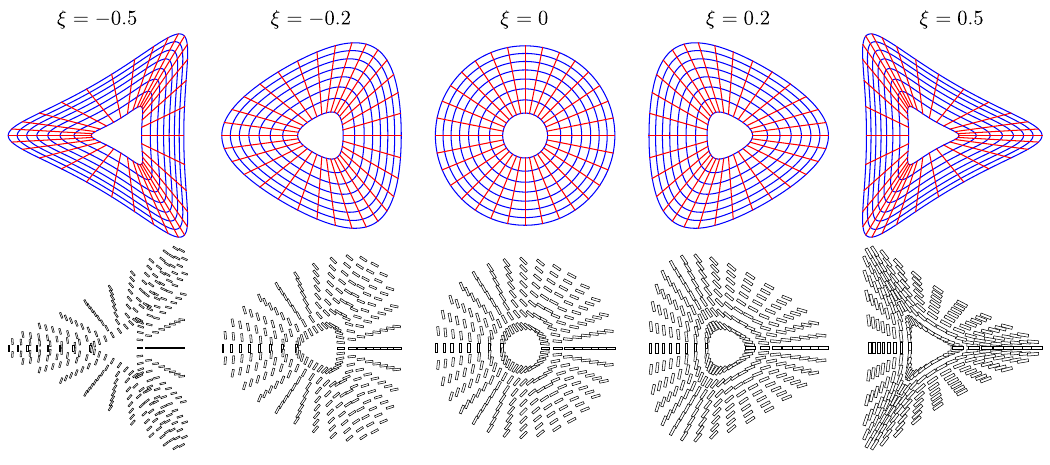}
\par\end{centering}
\caption{\textbf{Quasiconformal construction of the anisotropic $-\frac{1}{2}$
defect.} Projections of a polar grid on a reference anticone are shown
in the physical slit domain with corresponding director fields for
several values of the anisotropy parameter $\xi$. In the isotropic
limit, $\Omega_{-\frac{1}{2}}=3\pi$ and the complete construction
reduces to the conformal power map $z=w^{2/3}$. }\label{fig:DefectMaps2}
\end{figure*}

To solve the system Eqs. (\ref{eq:CompatibilityPsi}) and (\ref{eq:CompatibilitySystem}),
we begin by pulling the Dzyaloshinskii relation to the $c$-cone.
Using the relation between angular coordinates (\ref{eq:pullbackOfAngularCoordinate}),
and the definitions $p=-q$ and $\beta=\theta-\varphi$, we can transform
the Dzyaloshinskii solution to obtain,
\begin{align}
\psi & =-\Xi(\beta),\label{eq:PsiNewRelation}
\end{align}
which acquires an elegant geometric interpretation: it gives the angular
coordinate $\psi$ of the director configuration on the distorted
cone. On the chosen unwrapped $+\frac{1}{2}$ branch, $\theta$ increases
from $0$ to $\pi$ as $\varphi$ increases from $0$ to $2\pi$,
and hence $\beta=\theta-\varphi$ decreases from $0$ to $-\pi$.

The compatibility equations can now be integrated using Eq. (\ref{eq:PsiNewRelation}),
as detailed in Appendix A. Defining,
\begin{equation}
\mathcal{A}(\beta)=1+\epsilon\cos2\beta,\mathcal{B}(\beta)=1+q^{2}\epsilon\cos2\beta,\label{eq:ABbeta}
\end{equation}
together with, 
\begin{equation}
\mathcal{D}(\beta)=\sqrt{\mathcal{B}(\beta)}+q\sqrt{\mathcal{A}(\beta)},
\end{equation}
and the integral,
\begin{equation}
\mathcal{I}(\beta)=\int_{0}^{\beta}\frac{dt}{\sqrt{\mathcal{A}(t)\mathcal{B}(t)}},\label{eq:IIntegral}
\end{equation}
we obtain the angular parametrization,
\begin{equation}
\eta(\beta)=-\Omega\frac{\mathcal{I}(\beta)}{\mathcal{I}(\pi)},
\end{equation}
and the radial displacement,
\begin{equation}
h(\beta)=-\frac{1}{q}\log\left[\frac{D(\beta)}{D(0)}\right].
\end{equation}
Note that since $\mathcal{A}(\beta)$ and $\mathcal{B}(\beta)$ are
periodic with period $\pi$, the radial displacement satisfies $h(0)=h(-\pi)=0$
while the angular parametrization satisfies $\eta(0)=\psi(0)=0$ and
$\eta(-\pi)=\psi(-\pi)=\Omega$. Hence, the resulting map respects
the seam condition of the auxiliary cones and is globally compatible.

Collecting these expressions yields an explicit expression for the
quasiconformal map between the reference cone and the distorted cone,
\begin{equation}
H\left(Re^{i\eta(\beta)}\right)=R^{\kappa}e^{h(\beta)}e^{-i\Xi(\beta)}.\label{eq:THEMAP}
\end{equation}
where,
\begin{equation}
\kappa=\frac{\sqrt{1-\epsilon^{2}}}{\Omega}\mathcal{I}(\pi).
\end{equation}
Composing this with the power map $z=c^{q}$ gives the complete map
into the physical slit domain,
\begin{equation}
z\left(Re^{i\eta(\beta)}\right)=R^{q\kappa}\frac{\mathcal{D}(0)}{\mathcal{D}(\beta)}e^{-iq\Xi(\beta)}.\label{eq:ExplicitMap}
\end{equation}

We use these maps to visualize the image of the reference cone under
the quasiconformal map and the complete map to the physical domain
in Fig. \ref{fig:DefectMaps}, together with the corresponding director
profiles. Elastic anisotropy has two effects: it determines the cone
angle of the reference domain, and induces the quasiconformal distortion.
For $K_{3}<K_{1}$, where $\epsilon>0$, the defect is bend rich;
conversely for $\epsilon<0$ where $K_{3}>K_{1}$ the director profile
is splay rich; the bend or splay rich regions are quasiconformally
grown or shrunk as appropriate by the intermediate map.

The above construction can be readily adapted to other defect strengths
$s$, with the exception of $s=1$ for which the closure condition
Eq. (\ref{eq:closure}) is degenerate. During one counter-clockwise
circuit of the defect, $\Delta\theta=2\pi s$ and $\Delta\varphi=2\pi$,
so the relative director angle changes by $\beta_{s}=\Delta\beta=2\pi(s-1)$.
Hence, $\beta$ must run over the interval $[0,\beta_{s}]$. The cone
angle is $\Omega_{s}=\left|\Xi(\beta_{s})\right|=2\left|s-1\right|\Xi(\pi)$
and the closure condition uses the new cone angle $q_{s}\Omega_{s}=2\pi$
with $q_{s}=\left|p\right|$. The final conformal map must retain
the signed exponent,
\begin{equation}
z=c^{-p_{s}},\label{eq:PowerMap}
\end{equation}
so that $s<1$, for which $p_{s}<0$ gives a positive power and $s>1$,
for which $p_{s}>0$ gives a negative power map involving inversion.
In the quasiconformal construction, $\mathcal{I}(\pi)$ is replaced
by $\left|\mathcal{I}(\beta_{s})\right|$ so that, 
\begin{equation}
\kappa=\frac{\sqrt{1-\epsilon^{2}}}{\Omega_{s}}\left|\mathcal{I}(\beta_{s})\right|,
\end{equation}
and the angular coordinate becomes,
\begin{equation}
\eta(\beta)=-\Omega_{s}\frac{\mathcal{I}(\beta)}{\left|\mathcal{I}(\beta_{s})\right|},
\end{equation}
while the radial equation is unchanged. 

We display in Fig. \ref{fig:DefectMaps2} the image of the reference
cone (really an anti-cone as $\Omega_{-\frac{1}{2}}>2\pi$) mapped
into the physical domain for a $s=-1/2$ defect for various values
of $\xi$. Again, splay and bend rich regions are compressed or expanded
under the action of the quasiconformal map. Strikingly, the images
of constant radius curves on the reference cone in both Fig. \ref{fig:DefectMaps2}
and in the $+1/2$ case Fig. \ref{fig:DefectMaps} resemble the shapes
of the defect cores observed in the Q-tensor calculations, Ref. \citep{Schimming2020},
suggesting that the structure of these cores is attributable, at least
in part, to the underlying quasiconformal geometry. 

\subsection{Spin connection and defect holonomy}

We can examine the spin connection induced by the explicit map, Eq.
(\ref{eq:ExplicitMap}). From the phase of $c_{w}$, the rotation
associated with the quasiconformal map is,
\begin{equation}
\gamma_{H}=\psi-\eta+\delta(\beta),\ \delta(\beta)=\arg\left(1-\xi e^{i2\beta}\right).
\end{equation}
The conformal power map $z=c^{q}$ contributes an additional rotation
$(q-1)\psi$, so the total frame rotation and connection are,
\begin{equation}
\gamma=q\psi-\eta+\delta(\beta),\ \omega=d\gamma.
\end{equation}
The director angle in the rotating frame is,
\begin{equation}
\vartheta=\theta-\gamma=\beta+\eta-\delta(\beta),
\end{equation}
and hence $d\theta=d\vartheta+\omega$, recovering the frame covariant
form of the energy Eq. (\ref{eq:EnergyFrameCovariant}). 

For the $+1/2$ defect, a complete circuit corresponds to $\Delta\psi=\Delta\eta=\Omega$,
with $q\Omega=2\pi$. Since $|\xi|<1$, the function $\delta(\beta)$
has zero net winding, and hence the accumulated frame rotation is,
\begin{equation}
\Delta\gamma=q\Omega-\Omega=2\pi-\Omega.\label{eq:FrameRotation}
\end{equation}

Although the connection $\omega=d\gamma$ is locally pure gauge and
curvature-free away from the defect, the global seam identification
produces a $SO(2)$ holonomy $e^{i\Delta\gamma}=e^{i(2\pi-\Omega)}$,
corresponding to the deficit or excess angle of the auxiliary cone.
This parallels the geometric holonomy associated with liquid-crystal
order on physical conical substrates \citep{VafaZhangNelson2022,ZhangNelson2022}.
The quasiconformal map redistributes the local frame rotation but
contributes no additional net rotation between the equal-angle cones.

\subsection{The $s=+2$ defect}

\begin{figure*}
\includegraphics[width=1\textwidth]{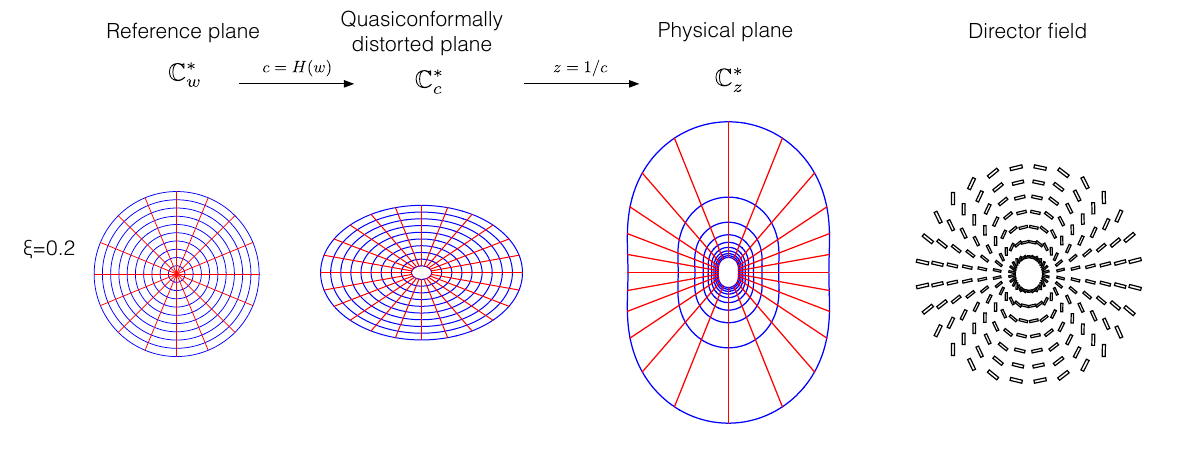}

\caption{\textbf{Quasiconformal geometry of the anisotropic $+2$ defect.}
Projections of a polar grid on a reference punctured plane are shown
on the quasiconformally stretched intermediate domain and in the physical
domain with a corresponding director field for $\xi=0.2$. In the
limit $\xi\to0$, the quasiconformal stretch becomes the identity
map and the defect is assembled by pure inversion.}\label{fig:PlusTwoMap}
\end{figure*}

The $s=+2$ defect is a special case where the quasiconformal mapping
derived previously simplifies greatly. The closure condition Eq. (\ref{eq:closure})
admits the exact solution $p=1$ independent of elastic anisotropy
because the numerator and denominator of the integrand cancel and
the integral evaluates to $\pi$. Around the defect, $\Delta\theta=2\pi s=4\pi$
and the relative director angle changes by $\beta_{s}=\Delta\beta=2\pi(s-1)=2\pi$. 

Examining the construction,
\begin{equation}
\mathcal{C}_{\Omega}\xrightarrow{c=H(w)}\mathcal{C}'_{\Omega}\xrightarrow{z=c^{-p}}\mathcal{S}_{z},\label{eq:TwoStepMap-1}
\end{equation}
we see that the conformal part of the map $z=c^{-p}=1/c$ becomes
pure inversion. Consistent with this, the reference cone has angle
$\Omega=2\pi$; it is simply the punctured plane $\mathbb{C}^{*}$.
Hence the sequence of domains for this special case simplifies to,
\begin{equation}
\mathbb{C}_{w}^{*}\xrightarrow{c=H(w)}\mathbb{C}_{c}^{*}\xrightarrow{z=c^{-p}}\mathbb{C}_{z}^{*},
\end{equation}
and both the quasiconformal and conformal maps are automorphisms.

We now derive the quasiconformal part. First, note that the director
in the physical plane has the simple expression, $\theta=2\varphi,$
and that inversion relates the coordinates $\mathcal{R}=r^{-1},\ \psi=-\varphi,$
so that $\theta=-2\psi$. We first pull the director back from the
physical domain to the $c$-plane, noting that an inversion $z=c^{-1}$
with $z_{c}=-c^{-2}$ rotates an arbitrary vector, 
\begin{align}
\arg z_{c} & =\pi-2\psi,\label{eq:RotationFromPowerMap-1}
\end{align}
so the director pulled back onto the $c$-plane is,
\begin{align}
\Theta & =\theta-\arg z_{c}=-\pi.
\end{align}
Since the director is defined modulo $\pi$, we may identify this
as $\Theta=0$. Hence, the $+2$ defect is transformed by inversion
into a uniform state in the $c$-plane. The compatibility condition,
\begin{equation}
c_{\bar{w}}=\xi e^{i2\Theta}\overline{c_{w}}=\xi\overline{c_{w}},\label{eq:Compatibility}
\end{equation}
now has a constant coefficient and corresponds to the uniform anisotropic
stretch introduced in Section \ref{sec:Quasiconformal-maps}. The
quasiconformal map is,
\begin{equation}
c=H(w)=\frac{w+\xi\bar{w}}{1+\xi},
\end{equation}
so that $c_{w}=(1+\xi)^{-1}$ and, $c_{\bar{w}}=\xi(1+\xi)^{-1}$,
which manifestly satisfy Eq. (\ref{eq:Compatibility}). The complete
$s=2$ map is therefore,
\begin{equation}
z(w)=\frac{1}{H(w)}=\frac{1+\xi}{w+\xi\bar{w}}.\label{eq:CompleteS2Map}
\end{equation}

In this special case, the quasiconformal map introduces anisotropic
stretching but \textit{does not} rotate the spatially uniform director
(in the $w$ plane), while the conformal inversion generates the entire
defect winding. Geometrically, this is consistent with the observation
that the defect texture for $s=+2$ does not depend on the LC anisotropy.
Nonetheless it \textit{does} possess nontrivial quasiconformal geometry:
as illustrated in Fig. \ref{fig:PlusTwoMap} for $\xi=0.2$, circles
in the reference plane acquire anisotropic shapes in the physical
domain, suggesting a geometric mechanism for the deformation of resolved
defect cores. We investigate this connection through numerical experiments
below.

\section{Numerical Experiments}\label{sec:Numerical-Experiments}

We now seek to assess the influence of the quasiconformal geometry
by examining the core structure of defects. We largely follow the
approach of Ref. \citep{Schimming2020}, which already clarified the
structure of $+\frac{1}{2}$ and $-\frac{1}{2}$ defects in the presence
of elastic anisotropy; calculations presented below permit us to examine
$+2$ defects. A similar approach was recently used to study the equilibrium
configuration of a disclination dipole\citep{Myers2024}.

Our simulations use the familiar $Q$-tensor order parameter to describe
the nematic phase, which is traceless and symmetric and related to
the nematic director $\mathbf{n}$ through the relation,
\begin{equation}
Q(\mathbf{x})=S(\mathbf{x})\left(\mathbf{n}(\mathbf{x})\otimes\mathbf{n}(\mathbf{x})-\frac{I_{3}}{3}\right),\label{eq:Qtensor}
\end{equation}
where $S$ is an additional scalar field describing the local degree
of alignment and $I_{3}$ is the identity matrix. We note that $Q$
is a 3D object embedded in two dimensions, allowing for phenomena
including escape from the plane, biaxiality, etc. The field $Q$ is
found by minimizing the energy,
\begin{equation}
F=\int_{C}f_{LdG}(Q)+f_{el}(Q,\nabla Q)d^{2}x,\label{eq:NumericalEnergy}
\end{equation}
where,
\begin{equation}
f_{LdG}(Q)=\frac{A}{2}\text{Tr}Q^{2}-\frac{B}{3}\text{Tr}Q^{3}+\frac{C}{4}\left(\text{Tr}Q^{2}\right)^{2},\label{eq:LDG}
\end{equation}
is the Landau-De Gennes energy that promotes the nematic phase, and,
\begin{align}
f_{el}(Q,\nabla Q)= & L_{1}\partial_{k}Q_{ij}\partial_{k}Q_{ij}+L_{2}\partial_{j}Q_{ij}\partial_{k}Q_{ik}\nonumber \\
 & +L_{3}Q_{kl}\partial_{k}Q_{ij}\partial_{l}Q_{ij},\label{eq:Qaniso}
\end{align}
is the elastic energy with elastic constants $L_{1}$, $L_{2}$ and
$L_{3}$. A number of different choices of prefactors and notation
exist in the literature for Eqs. (\ref{eq:LDG}) and (\ref{eq:Qaniso});
Ref. \citep{MottramNewton2014} refers to $L_{3}$ from the present
as $L_{6}$, for example. Note that here we use the Landau-De Gennes
functional for simplicity; Ref. \citep{Schimming2020} uses an alternative
functional with significantly improved stability properties. 

The elastic constants are related to the Frank constants,
\begin{align}
K_{1} & =2S_{0}^{2}\left(2L_{1}+L_{2}\right)-\frac{4}{3}L_{3}S_{0}^{3},\label{eq:FrankToQK1}\\
K_{3} & =2S_{0}^{2}\left(2L_{1}+L_{2}\right)+\frac{8}{3}L_{3}S_{0}^{3},\label{eq:FrankToQK3}
\end{align}
under the assumption that the scalar order parameter $S$ takes the
spatially uniform value $S_{0}$. Combining Equations (\ref{eq:FrankToQK1})
and (\ref{eq:FrankToQK3}) yields,
\begin{equation}
K_{3}-K_{1}=4L_{3}S_{0}^{3},\label{eq:K3K1isL3}
\end{equation}
so we see the parameter $L_{3}$ sets the degree of elastic anisotropy.

We will minimize Eq. \ref{eq:NumericalEnergy} with respect to suitable
parametrizations of $Q$ and subject to Dirichlet conditions on the
boundary of a square domain. We examine two sets of parameter values
for the elastic constants $L_{1}$ and $L_{2}$, varying $L_{3}$
as motivated by Eq. (\ref{eq:K3K1isL3}). The first follows Ref. \citep{Schimming2020}
with $L_{1}=1,L_{2}=7$; we also explored $L_{1}=4.5$, $L_{2}=0$,
preserving the value of $2L_{1}+L_{2}$ and hence yielding identical
values of $K_{1}$ and $K_{3}$ for given $L_{3}$. For the Landau
terms, we choose $A=29.16864$, $B=648.096$, $C=384$ that promote
an equilibrium value of $S_{0}=0.6751$, matching the value in Ref.
\citep{Schimming2020}. These choices set a natural defect length
scale $l_{core}\sim\sqrt{\left\langle L\right\rangle /C/S_{0}^{2}}\approx0.2$
in simulation units, although the defect itself can be significantly
larger as it depends on the amount of elastic distortion; we use a
simulation domain of side $\lambda=20$ that is much greater than
the core size. 
\begin{figure}
\includegraphics[width=1\columnwidth]{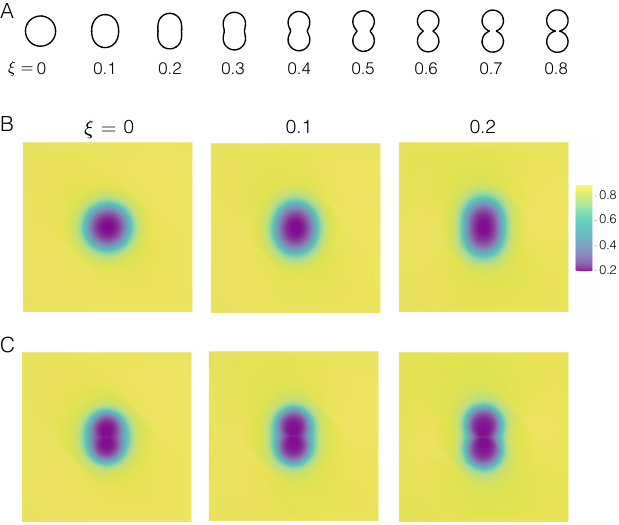}

\caption{\textbf{Core structure of a $+2$ defect.} \textbf{A} Projection
of a circle in the reference plane to the physical plane under the
action of the quasiconformal map Eq. (\ref{eq:CompleteS2Map}); the
shapes are scaled to have equal enclosed area.\textbf{ }Minimized
scalar order parameters for fixed director field and \textbf{B} elastic
constants $L_{1}=4.5$, $L_{2}=0$ and \textbf{C} $L_{1}=1$, $L_{2}=7$;
in both cases $L_{3}$ is varied to give $\xi=[0,0.1,0.2]$. }\label{fig:CorePlus2}

\end{figure}

Since a $s=+2$ defect is generally unstable against decomposition
into other defects, to study the structure we perform a constrained
minimization: we construct the $Q$ tensor Eq. (\ref{eq:Qtensor})
with a fixed director field using $\theta=2\arg(x+iy)$, a known minimizer
of the Frank energy independent of $K_{1}$ and $K_{3}$. We then
minimize Eq. (\ref{eq:NumericalEnergy}) with respect to the scalar
order parameter $S(\mathbf{x})$ only. Cursor, using Grok 4.7, was
employed to develop prototype simulation scripts in \textit{Morpho}\citep{Joshi2025Morpho};
these were subsequently reviewed and refined by the author, and the
final programs are provided as Supplemental Material.

In Fig. \ref{fig:CorePlus2}A, we show how a circular contour is deformed
under the map Eq. (\ref{eq:CompleteS2Map}) for values of anisotropy
$\xi\in[0,0.8]$. We then show minimized $S$ for $L_{1}=4.5$, $L_{2}=0$
in Fig. \ref{fig:CorePlus2}B and $L_{1}=1$, $L_{2}=7$ in Fig. \ref{fig:CorePlus2}C.
In both cases, the core morphologies closely resemble members of the
family of shapes generated by the quasiconformal construction. For
the first case with $L_{2}=0$, the shapes match at the same values
of $\xi$. 

In the second case, the cores are strongly anisotropic even at $\xi=0$,
where the Frank constants are equal. The anisotropy arises from the
$L_{2}$ term, which couples gradients of $S$ to the director and
has no counterpart in the Frank energy. Nonetheless, the resulting
core shapes recognizably remain members of the quasiconformal family,
corresponding to an effective anisotropy $\xi_{eff}$ larger than
that implied by $K_{1}$ and $K_{3}$. Hence, the geometric structure
predicted by the quasiconformal map appears to persist in defect cores
even when the physical mechanism generating the anisotropy is beyond
the Frank theory.

\section{Discussion}\label{sec:Discussion}

We conclude by considering two connections to other work: the measurement
of elastic anisotropy from defect textures and the geometric description
of programmed deformations in liquid crystal elastomers.

\subsection{Measuring elastic anisotropy}

Since the configuration of $\mathbf{n}$ around a defect typically
depends strongly on elastic anisotropy, it provides a natural route
to measure it\citep{hudson1989frank}. Multiplying Eq. (\ref{eq:Frank2D})
by a common constant does not change the equilibrium solution, so
it is not possible to determine $K_{1}$ and $K_{3}$ simultaneously.
The texture may, however, determine their ratio $K_{3}/K_{1}$ or
equivalently the parameters $\xi$ and $\epsilon$ from above; we
therefore connect our work here with another recent approach.

Klein and Beller\citep{KleinBeller2026} have recently proposed a
family of order parameters to measure the elastic anisotropy from
defect textures,

\begin{equation}
\Gamma(R)=\frac{1}{2\pi R}\oint\left(\mathbf{n}\cdot\hat{\mathbf{r}}\right)^{2}dl=\left\langle \left(\mathbf{n}\cdot\hat{\mathbf{r}}\right)^{2}\right\rangle _{R},
\end{equation}
where $\hat{\mathbf{r}}$ is the radial unit vector from the defect
core, the integral is to be evaluated over a loop of radius $R$,
and the notation $\left\langle f\right\rangle _{R}$ defines a similar
contour moment taken at a specified radius. Using the relative director
angle $\beta=\theta-\varphi$, we have $2\Gamma-1=\langle\cos2\beta\rangle_{R}$,
which measures the preference for radial or tangential director alignment.

This quantity has a natural interpretation in the quasiconformal framework.
The inverse isotropizing map $w(z)$ has Beltrami coefficient,
\begin{equation}
\nu(z)=\frac{w_{\bar{z}}}{w_{z}}=-\xi e^{2i\theta(z)},
\end{equation}
and hence $e^{-2i\varphi}\nu=-\xi e^{2i\beta}$. Thus, the angular
quantity measured by $\Gamma$ is directly related to the orientation
of the quasiconformal distortion relative to the radial frame.

For an isolated equilibrium $+1/2$ defect, the Dzyaloshinskii solution
gives, to leading order in elastic anisotropy,
\[
2\Gamma-1\approx\frac{3}{2}\xi+O(\xi^{3}).
\]
The angular structure of the defect therefore provides a direct measure
of the signed quasiconformal anisotropy parameter $\xi$ in this limit.

\subsection{Connection with Liquid Crystal Elastomers}

We also consider applications of quasiconformal geometry to a closely
related system, liquid crystal elastomers. The polymer step-length
tensor of the neo-classical theory of liquid-crystal elastomers,
\begin{equation}
\mathit{l}=l_{\perp}I+(l_{\parallel}-l_{\perp})\mathbf{n}\otimes\mathbf{n},
\end{equation}
is algebraically identical to the elastic anisotropy tensor in Eq.
(\ref{eq:lcelasticmetric}). In the Frank problem, $A$ weights gradients,
while in an elastomer $l$ encodes the preferred spontaneous deformation
and a target metric\citep{Aharoni2014}. This suggests a parallel
between elastic anisotropy in regular liquid crystals and liquid crystal
elastomers: here we found a transformation to an auxiliary coordinate
system that rendered the quadratic form isotropic; in an LCE the material
preparation effectively records a preferred coordinate system, and
the material changes shape attempting to achieve it. Elegant geometric
work has shown that radial and azimuthal director patterns in LCEs
produce cones, anticones, saddles, etc. upon heating or cooling\citep{Modes2010,Modes2011Gaussian,Aharoni2014,Efrati2009}.

Our results provide a possible new perspective on these transformations:
such cones may be thought of as the reference geometries of a quasiconformal
construction made manifest in a real surface. Following the parallel,
Beltrami coefficient $\mu$ in an LCE encodes the local spontaneous
strain, and hence the experimental patterning can be thought of as
specifying local values of $\mu$. Such a connection with Beltrami
equations was recognized by Aharoni, Sharon and Kupferman in their
treatment of the inverse metric problem for thin nematic elastomer
sheets\citep{Aharoni2014}. Our results here suggest that quasiconformal
maps may serve more broadly as an organizing language connecting director
elasticity, programmed metrics and defect geometry. Speculatively,
liquid crystals and liquid crystal elastomers may prove to be two
realizations of the same underlying quasiconformal geometry: one is
encoded in the energy, the other in the material itself.

\section{Conclusions}\label{sec:Conclusions}

We have shown that elastic anisotropy in a liquid crystal defines
a quasiconformal geometry whose principal stretches are fixed by the
elastic moduli and whose axes rotate with the director; the geometry
is not prescribed but emerges alongside the director as the solution
of an underlying variational problem. Local compatibility requires
an accompanying local dilation and frame rotation, naturally introducing
a spin connection that may lead to nontrivial holonomy in suitable
geometries. 

Applying this construction reveals new insight into anisotropic liquid
crystal textures: a hybrid-aligned film is associated with an explicit
shear map that isotropizes the complete elastic tensor. For isolated
defects, Dzyaloshinskii’s closure condition acquires a new geometric
interpretation: it fixes the total angle of an auxiliary reference
cone. The anisotropic defect texture is then generated by a quasiconformal
map between equal-angle cones followed by a conformal power map onto
the physical slit domain. In the one constant limit, the quasiconformal
part becomes the identity and the construction reduces to the classical
conformal construction.

Numerical simulations of $+2$ defects, together with existing results
in the literature, demonstrate that the quasiconformal construction
appears to describe defect core morphology. For the $+2$ defects,
the resulting core shapes remain members of the same quasiconformal
family even when the mechanism producing the anisotropy is beyond
the Frank-Oseen theory alone. The geometry thus provides not only
an interpretation of the observed core structures, but also a means
of probing physical effects absent in the director-only theory.

These results provide a geometric interpretation of elastic anisotropy,
a geometrically motivated anisotropy parameter $\xi=\left(\sqrt{K_{3}}-\sqrt{K_{1}}\right)/\left(\sqrt{K_{3}}+\sqrt{K_{1}}\right)$,
and suggest a route for identifying analogous quasiconformal structures
in more general liquid-crystal configurations. The framework is general
and naturally suggests a number of avenues for further work: Configurations
containing multiple defects naturally lead from the cone geometry
of the isolated case to the quasiconformal geometry of marked and
punctured surfaces. In this setting, defect positions, boundary data
and seam identifications become moduli, suggesting a connection with
Teichmüller theory and with the extremal quasiconformal maps generated
by quadratic differentials. It would be interesting to determine whether
liquid-crystal equilibrium selects distinguished representatives within
these quasiconformal classes and how or whether the holonomy of the
induced frame connection encodes additional information such as interactions
among defects. 

Beyond two dimensions, a single Beltrami coefficient no longer exists
and the analogous problem involves spatially varying elastic metrics
together with a more general geometry of local frames. As an intermediate
step, it is natural to ask what becomes of the quasiconformal structure
even in two dimensions when the director is allowed to escape from
the plane; the results may be relevant to the interpretation of disclination
lines, escaped textures, hedgehogs and three-dimensional topological
solitons.

Finally, extending the construction to time-dependent and active textures
would promote the Beltrami coefficient and frame connection to evolving
geometrical fields, providing a possible language for defect dynamics
in anisotropic media. Indeed, quasiconformal geometry may provide
a unifying language in two dimensions for other kinds of anisotropy
in systems beyond liquid crystals as hinted in previous work in optics,
conduction problems and morphogenesis. 
\begin{acknowledgments}
The author is grateful to Brandon Klein and Daniel Beller for valuable
conversations. \textit{Statement on AI use:} The author acknowledges
the use of OpenAI's ChatGPT (GPT-6) to assist with exploratory calculations,
manuscript editing, reviewing mathematical derivations for consistency,
and discussing the interpretation of results. The author independently
verified mathematical results using Mathematica and hand calculations. 
\end{acknowledgments}

\section*{Appendix A}

In this Appendix, we use the compatibility equations Eqs. (\ref{eq:CompatibilityPsi})
and (\ref{eq:CompatibilitySystem}) together with Eq. (\ref{eq:PsiNewRelation})
to derive the angular parametrization and radial displacement that
define the quasiconformal map $c=H(w)$ for anisotropic defects. 

The numerator and denominator of the Dzyaloshinskii integral are defined
as the two functions $\mathcal{A}(\beta)=1+\epsilon\cos2\beta$ and
$\mathcal{B}(\beta)=1+q^{2}\epsilon\cos2\beta$ as above. Differentiating
the relation $\psi=-\Xi(\beta)$, Eq. (\ref{eq:DzyaloshinskiiAngle}),
gives,
\begin{equation}
\psi_{\beta}=-\sqrt{\mathcal{A}(\beta)/\mathcal{B}(\beta)}.\label{eq:DzyaloshinskiiIntegrand}
\end{equation}
With these definitions, the first compatibility equation Eq. (\ref{eq:CompatibilityPsi})
can be written,
\begin{equation}
\psi_{\eta}=\frac{\kappa}{\sqrt{1-\epsilon^{2}}}\mathcal{A}(\beta),
\end{equation}
which, using the relation $\psi_{\eta}=\psi_{\beta}\beta_{\eta}$
can be transformed into,
\begin{equation}
\beta_{\eta}=-\frac{\kappa}{\sqrt{1-\epsilon^{2}}}\sqrt{\mathcal{A}(\beta)\mathcal{B}(\beta)}.
\end{equation}
Inverting the derivative yields,
\begin{equation}
\eta_{\beta}=-\frac{\sqrt{1-\epsilon^{2}}}{\kappa}\frac{1}{\sqrt{\mathcal{A}(\beta)\mathcal{B}(\beta)}}.\label{eq:eta_beta}
\end{equation}
The integral of this quantity as in Eq. (\ref{eq:IIntegral}) can
be expressed in terms of elliptic functions for calculation purposes.
Choosing $\eta=0$ at $\beta=0$ and integrating Eq. (\ref{eq:eta_beta})
gives,
\begin{equation}
\eta(\beta)=-\frac{\sqrt{1-\epsilon^{2}}}{\kappa}\mathcal{I}(\beta).\label{eq:etaRelation}
\end{equation}
The constant $\kappa$ is obtained using $\eta=\Omega$ when $\beta=-\pi$,
\begin{equation}
\kappa=\frac{\sqrt{1-\epsilon^{2}}}{\Omega}\mathcal{I}(\pi).\label{eq:kapparelation}
\end{equation}
Combining Eq. (\ref{eq:etaRelation}) and Eq. (\ref{eq:kapparelation}),
together with Eq. (\ref{eq:PsiNewRelation}), the angular part of
the quasiconformal map can therefore be expressed parametrically,
\begin{equation}
\eta(\beta)=-\Omega\frac{\mathcal{I}(\beta)}{\mathcal{I}(\pi)},\ \psi(\beta)=-\Xi(\beta),\ -\pi\le\beta\le0.\label{eq:AngleQCRESULT}
\end{equation}

The second compatibility equation Eq. (\ref{eq:CompatibilitySystem})
can be similarly transformed,
\begin{equation}
h_{\beta}=\frac{\epsilon\sin2\beta}{\sqrt{\mathcal{A}(\beta)\mathcal{B}(\beta)}}.
\end{equation}
Letting $h(0)=0$, and using the definition $\mathcal{D}(\beta)=\sqrt{\mathcal{B}(\beta)}+q\sqrt{\mathcal{A}(\beta)}$
and the identity,
\begin{equation}
\frac{d}{d\beta}\log\mathcal{D}(\beta)=-\frac{q\epsilon\sin2\beta}{\sqrt{\mathcal{A}(\beta)\mathcal{B}(\beta)}},
\end{equation}
the integral,
\begin{equation}
h(\beta)=\int_{0}^{\beta}\frac{\epsilon\sin2t}{\sqrt{\mathcal{A}(t)\mathcal{B}(t)}}dt,
\end{equation}
can be evaluated in closed form,
\begin{equation}
h(\beta)=-\frac{1}{q}\log\left[\frac{\mathcal{D}(\beta)}{\mathcal{D}(0)}\right].\label{eq:RadialQCRESULT}
\end{equation}
Equations (\ref{eq:AngleQCRESULT}) and (\ref{eq:RadialQCRESULT}),
representing the angular parametrization and radial displacements
respectively, can be composed to form the full quasiconformal map
as described in the main text. \bibliographystyle{apsrev4-2}

\begin{thebibliography}{40}%
\makeatletter
\providecommand \@ifxundefined [1]{%
 \@ifx{#1\undefined}
}%
\providecommand \@ifnum [1]{%
 \ifnum #1\expandafter \@firstoftwo
 \else \expandafter \@secondoftwo
 \fi
}%
\providecommand \@ifx [1]{%
 \ifx #1\expandafter \@firstoftwo
 \else \expandafter \@secondoftwo
 \fi
}%
\providecommand \natexlab [1]{#1}%
\providecommand \enquote  [1]{``#1''}%
\providecommand \bibnamefont  [1]{#1}%
\providecommand \bibfnamefont [1]{#1}%
\providecommand \citenamefont [1]{#1}%
\providecommand \href@noop [0]{\@secondoftwo}%
\providecommand \href [0]{\begingroup \@sanitize@url \@href}%
\providecommand \@href[1]{\@@startlink{#1}\@@href}%
\providecommand \@@href[1]{\endgroup#1\@@endlink}%
\providecommand \@sanitize@url [0]{\catcode `\\12\catcode `\$12\catcode
  `\&12\catcode `\#12\catcode `\^12\catcode `\_12\catcode `\%12\relax}%
\providecommand \@@startlink[1]{}%
\providecommand \@@endlink[0]{}%
\providecommand \url  [0]{\begingroup\@sanitize@url \@url }%
\providecommand \@url [1]{\endgroup\@href {#1}{\urlprefix }}%
\providecommand \urlprefix  [0]{URL }%
\providecommand \Eprint [0]{\href }%
\providecommand \doibase [0]{https://doi.org/}%
\providecommand \selectlanguage [0]{\@gobble}%
\providecommand \bibinfo  [0]{\@secondoftwo}%
\providecommand \bibfield  [0]{\@secondoftwo}%
\providecommand \translation [1]{[#1]}%
\providecommand \BibitemOpen [0]{}%
\providecommand \bibitemStop [0]{}%
\providecommand \bibitemNoStop [0]{.\EOS\space}%
\providecommand \EOS [0]{\spacefactor3000\relax}%
\providecommand \BibitemShut  [1]{\csname bibitem#1\endcsname}%
\let\auto@bib@innerbib\@empty
\bibitem [{\citenamefont {Selinger}(2022)}]{Selinger2022}%
  \BibitemOpen
  \bibfield  {author} {\bibinfo {author} {\bibfnamefont {J.~V.}\ \bibnamefont
  {Selinger}},\ }\href
  {https://doi.org/10.1146/annurev-conmatphys-031620-105712} {\bibfield
  {journal} {\bibinfo  {journal} {Annual Review of Condensed Matter Physics}\
  }\textbf {\bibinfo {volume} {13}},\ \bibinfo {pages} {49} (\bibinfo {year}
  {2022})}\BibitemShut {NoStop}%
\bibitem [{\citenamefont {Jeong}\ \emph {et~al.}(2014)\citenamefont {Jeong},
  \citenamefont {Davidson}, \citenamefont {Collings}, \citenamefont
  {Lubensky},\ and\ \citenamefont {Yodh}}]{Jeong2014}%
  \BibitemOpen
  \bibfield  {author} {\bibinfo {author} {\bibfnamefont {J.}~\bibnamefont
  {Jeong}}, \bibinfo {author} {\bibfnamefont {Z.~S.}\ \bibnamefont {Davidson}},
  \bibinfo {author} {\bibfnamefont {P.~J.}\ \bibnamefont {Collings}}, \bibinfo
  {author} {\bibfnamefont {T.~C.}\ \bibnamefont {Lubensky}},\ and\ \bibinfo
  {author} {\bibfnamefont {A.~G.}\ \bibnamefont {Yodh}},\ }\href
  {https://doi.org/10.1073/pnas.1315121111} {\bibfield  {journal} {\bibinfo
  {journal} {Proceedings of the National Academy of Sciences}\ }\textbf
  {\bibinfo {volume} {111}},\ \bibinfo {pages} {1742} (\bibinfo {year}
  {2014})}\BibitemShut {NoStop}%
\bibitem [{\citenamefont {Schimming}\ and\ \citenamefont
  {Vi\~nals}(2020)}]{Schimming2020}%
  \BibitemOpen
  \bibfield  {author} {\bibinfo {author} {\bibfnamefont {C.~D.}\ \bibnamefont
  {Schimming}}\ and\ \bibinfo {author} {\bibfnamefont {J.}~\bibnamefont
  {Vi\~nals}},\ }\href {https://doi.org/10.1103/PhysRevE.102.010701} {\bibfield
   {journal} {\bibinfo  {journal} {Phys. Rev. E}\ }\textbf {\bibinfo {volume}
  {102}},\ \bibinfo {pages} {010701(R)} (\bibinfo {year} {2020})}\BibitemShut
  {NoStop}%
\bibitem [{\citenamefont {DeBenedictis}\ and\ \citenamefont
  {Atherton}(2016)}]{DeBenedictisAtherton2016}%
  \BibitemOpen
  \bibfield  {author} {\bibinfo {author} {\bibfnamefont {A.}~\bibnamefont
  {DeBenedictis}}\ and\ \bibinfo {author} {\bibfnamefont {T.~J.}\ \bibnamefont
  {Atherton}},\ }\href {https://doi.org/10.1080/02678292.2016.1209699}
  {\bibfield  {journal} {\bibinfo  {journal} {Liquid Crystals}\ }\textbf
  {\bibinfo {volume} {43}},\ \bibinfo {pages} {2352} (\bibinfo {year}
  {2016})}\BibitemShut {NoStop}%
\bibitem [{\citenamefont {Schimming}\ and\ \citenamefont
  {Vi{\~n}als}(2022)}]{SchimmingVinals2022}%
  \BibitemOpen
  \bibfield  {author} {\bibinfo {author} {\bibfnamefont {C.~D.}\ \bibnamefont
  {Schimming}}\ and\ \bibinfo {author} {\bibfnamefont {J.}~\bibnamefont
  {Vi{\~n}als}},\ }\href {https://doi.org/10.1039/D2SM00323F} {\bibfield
  {journal} {\bibinfo  {journal} {Soft Matter}\ }\textbf {\bibinfo {volume}
  {18}},\ \bibinfo {pages} {8024} (\bibinfo {year} {2022})}\BibitemShut
  {NoStop}%
\bibitem [{\citenamefont {Houston}(2026)}]{Houston2026}%
  \BibitemOpen
  \bibfield  {author} {\bibinfo {author} {\bibfnamefont {A.~J.~H.}\
  \bibnamefont {Houston}},\ }\href {https://doi.org/10.1098/rspa.2025.1033}
  {\bibfield  {journal} {\bibinfo  {journal} {Proceedings of the Royal Society
  A}\ }\textbf {\bibinfo {volume} {482}},\ \bibinfo {pages} {20251033}
  (\bibinfo {year} {2026})}\BibitemShut {NoStop}%
\bibitem [{\citenamefont {Koizumi}\ \emph {et~al.}(2023)\citenamefont
  {Koizumi}, \citenamefont {Golovaty}, \citenamefont {Alqarni}, \citenamefont
  {Li}, \citenamefont {Sternberg},\ and\ \citenamefont
  {Lavrentovich}}]{Koizumi2023}%
  \BibitemOpen
  \bibfield  {author} {\bibinfo {author} {\bibfnamefont {R.}~\bibnamefont
  {Koizumi}}, \bibinfo {author} {\bibfnamefont {D.}~\bibnamefont {Golovaty}},
  \bibinfo {author} {\bibfnamefont {A.}~\bibnamefont {Alqarni}}, \bibinfo
  {author} {\bibfnamefont {B.-X.}\ \bibnamefont {Li}}, \bibinfo {author}
  {\bibfnamefont {P.~J.}\ \bibnamefont {Sternberg}},\ and\ \bibinfo {author}
  {\bibfnamefont {O.~D.}\ \bibnamefont {Lavrentovich}},\ }\href
  {https://doi.org/10.1126/sciadv.adf3385} {\bibfield  {journal} {\bibinfo
  {journal} {Science Advances}\ }\textbf {\bibinfo {volume} {9}},\ \bibinfo
  {pages} {eadf3385} (\bibinfo {year} {2023})}\BibitemShut {NoStop}%
\bibitem [{\citenamefont {Atherton}\ and\ \citenamefont
  {Sambles}(2006)}]{AthertonSambles2006}%
  \BibitemOpen
  \bibfield  {author} {\bibinfo {author} {\bibfnamefont {T.~J.}\ \bibnamefont
  {Atherton}}\ and\ \bibinfo {author} {\bibfnamefont {J.~R.}\ \bibnamefont
  {Sambles}},\ }\href {https://doi.org/10.1103/PhysRevE.74.022701} {\bibfield
  {journal} {\bibinfo  {journal} {Physical Review E}\ }\textbf {\bibinfo
  {volume} {74}},\ \bibinfo {pages} {022701} (\bibinfo {year}
  {2006})}\BibitemShut {NoStop}%
\bibitem [{\citenamefont {Anquetil-Deck}\ \emph {et~al.}(2012)\citenamefont
  {Anquetil-Deck}, \citenamefont {Cleaver},\ and\ \citenamefont
  {Atherton}}]{AnquetilDeck2012}%
  \BibitemOpen
  \bibfield  {author} {\bibinfo {author} {\bibfnamefont {C.}~\bibnamefont
  {Anquetil-Deck}}, \bibinfo {author} {\bibfnamefont {D.~J.}\ \bibnamefont
  {Cleaver}},\ and\ \bibinfo {author} {\bibfnamefont {T.~J.}\ \bibnamefont
  {Atherton}},\ }\href {https://doi.org/10.1103/PhysRevE.86.041707} {\bibfield
  {journal} {\bibinfo  {journal} {Physical Review E}\ }\textbf {\bibinfo
  {volume} {86}},\ \bibinfo {pages} {041707} (\bibinfo {year}
  {2012})}\BibitemShut {NoStop}%
\bibitem [{\citenamefont {Anquetil-Deck}\ \emph {et~al.}(2013)\citenamefont
  {Anquetil-Deck}, \citenamefont {Cleaver}, \citenamefont {Bramble},\ and\
  \citenamefont {Atherton}}]{AnquetilDeck2013}%
  \BibitemOpen
  \bibfield  {author} {\bibinfo {author} {\bibfnamefont {C.}~\bibnamefont
  {Anquetil-Deck}}, \bibinfo {author} {\bibfnamefont {D.~J.}\ \bibnamefont
  {Cleaver}}, \bibinfo {author} {\bibfnamefont {J.~P.}\ \bibnamefont
  {Bramble}},\ and\ \bibinfo {author} {\bibfnamefont {T.~J.}\ \bibnamefont
  {Atherton}},\ }\href {https://doi.org/10.1103/PhysRevE.88.012501} {\bibfield
  {journal} {\bibinfo  {journal} {Physical Review E}\ }\textbf {\bibinfo
  {volume} {88}},\ \bibinfo {pages} {012501} (\bibinfo {year}
  {2013})}\BibitemShut {NoStop}%
\bibitem [{\citenamefont {DeBenedictis}\ \emph {et~al.}(2015)\citenamefont
  {DeBenedictis}, \citenamefont {Atherton}, \citenamefont {Anquetil-Deck},
  \citenamefont {Cleaver}, \citenamefont {Emerson}, \citenamefont {Wolak},\
  and\ \citenamefont {Adler}}]{DeBenedictis2015}%
  \BibitemOpen
  \bibfield  {author} {\bibinfo {author} {\bibfnamefont {A.}~\bibnamefont
  {DeBenedictis}}, \bibinfo {author} {\bibfnamefont {T.~J.}\ \bibnamefont
  {Atherton}}, \bibinfo {author} {\bibfnamefont {C.}~\bibnamefont
  {Anquetil-Deck}}, \bibinfo {author} {\bibfnamefont {D.~J.}\ \bibnamefont
  {Cleaver}}, \bibinfo {author} {\bibfnamefont {D.~B.}\ \bibnamefont
  {Emerson}}, \bibinfo {author} {\bibfnamefont {M.}~\bibnamefont {Wolak}},\
  and\ \bibinfo {author} {\bibfnamefont {J.~H.}\ \bibnamefont {Adler}},\ }\href
  {https://doi.org/10.1103/PhysRevE.92.042501} {\bibfield  {journal} {\bibinfo
  {journal} {Physical Review E}\ }\textbf {\bibinfo {volume} {92}},\ \bibinfo
  {pages} {042501} (\bibinfo {year} {2015})}\BibitemShut {NoStop}%
\bibitem [{\citenamefont {Chandler}\ and\ \citenamefont
  {Spagnolie}(2023)}]{ChandlerSpagnolie2023}%
  \BibitemOpen
  \bibfield  {author} {\bibinfo {author} {\bibfnamefont {T.~G.~J.}\
  \bibnamefont {Chandler}}\ and\ \bibinfo {author} {\bibfnamefont {S.~E.}\
  \bibnamefont {Spagnolie}},\ }\href {https://doi.org/10.1017/jfm.2023.488}
  {\bibfield  {journal} {\bibinfo  {journal} {Journal of Fluid Mechanics}\
  }\textbf {\bibinfo {volume} {967}},\ \bibinfo {pages} {A19} (\bibinfo {year}
  {2023})}\BibitemShut {NoStop}%
\bibitem [{\citenamefont {Chandler}\ and\ \citenamefont
  {Spagnolie}(2024)}]{ChandlerSpagnolie2024Interactions}%
  \BibitemOpen
  \bibfield  {author} {\bibinfo {author} {\bibfnamefont {T.~G.~J.}\
  \bibnamefont {Chandler}}\ and\ \bibinfo {author} {\bibfnamefont {S.~E.}\
  \bibnamefont {Spagnolie}},\ }\href {https://doi.org/10.1137/23M1622908}
  {\bibfield  {journal} {\bibinfo  {journal} {SIAM Journal on Applied
  Mathematics}\ }\textbf {\bibinfo {volume} {84}},\ \bibinfo {pages} {2476}
  (\bibinfo {year} {2024})}\BibitemShut {NoStop}%
\bibitem [{\citenamefont {Chandler}\ and\ \citenamefont
  {Spagnolie}(2025)}]{Chandler2025}%
  \BibitemOpen
  \bibfield  {author} {\bibinfo {author} {\bibfnamefont {T.~G.~J.}\
  \bibnamefont {Chandler}}\ and\ \bibinfo {author} {\bibfnamefont {S.~E.}\
  \bibnamefont {Spagnolie}},\ }\href {https://doi.org/10.1103/x8y2-tw4c}
  {\bibfield  {journal} {\bibinfo  {journal} {Phys. Rev. Res.}\ }\textbf
  {\bibinfo {volume} {7}},\ \bibinfo {pages} {023263} (\bibinfo {year}
  {2025})}\BibitemShut {NoStop}%
\bibitem [{\citenamefont {Needham}\ and\ \citenamefont
  {Penrose}(2023)}]{needham2023visual}%
  \BibitemOpen
  \bibfield  {author} {\bibinfo {author} {\bibfnamefont {T.}~\bibnamefont
  {Needham}}\ and\ \bibinfo {author} {\bibfnamefont {R.}~\bibnamefont
  {Penrose}},\ }\href {https://books.google.com/books?id=GoCsEAAAQBAJ} {\emph
  {\bibinfo {title} {Visual Complex Analysis: 25th Anniversary Edition}}}\
  (\bibinfo  {publisher} {OUP Oxford},\ \bibinfo {year} {2023})\BibitemShut
  {NoStop}%
\bibitem [{\citenamefont {Ahlfors}(2006)}]{Ahlfors2006}%
  \BibitemOpen
  \bibfield  {author} {\bibinfo {author} {\bibfnamefont {L.~V.}\ \bibnamefont
  {Ahlfors}},\ }\href {https://doi.org/10.1090/ulect/038} {\emph {\bibinfo
  {title} {Lectures on Quasiconformal Mappings}}},\ \bibinfo {edition} {2nd}\
  ed.,\ \bibinfo {series} {University Lecture Series}, Vol.~\bibinfo {volume}
  {38}\ (\bibinfo  {publisher} {American Mathematical Society},\ \bibinfo
  {address} {Providence, RI},\ \bibinfo {year} {2006})\BibitemShut {NoStop}%
\bibitem [{\citenamefont {Astala}\ \emph {et~al.}(2009)\citenamefont {Astala},
  \citenamefont {Iwaniec},\ and\ \citenamefont
  {Martin}}]{AstalaIwaniecMartin2009}%
  \BibitemOpen
  \bibfield  {author} {\bibinfo {author} {\bibfnamefont {K.}~\bibnamefont
  {Astala}}, \bibinfo {author} {\bibfnamefont {T.}~\bibnamefont {Iwaniec}},\
  and\ \bibinfo {author} {\bibfnamefont {G.}~\bibnamefont {Martin}},\
  }\href@noop {} {\emph {\bibinfo {title} {Elliptic Partial Differential
  Equations and Quasiconformal Mappings in the Plane}}},\ \bibinfo {series}
  {Princeton Mathematical Series}, Vol.~\bibinfo {volume} {48}\ (\bibinfo
  {publisher} {Princeton University Press},\ \bibinfo {address} {Princeton,
  NJ},\ \bibinfo {year} {2009})\BibitemShut {NoStop}%
\bibitem [{\citenamefont {Ahlfors}\ and\ \citenamefont
  {Bers}(1960)}]{AhlforsBers1960}%
  \BibitemOpen
  \bibfield  {author} {\bibinfo {author} {\bibfnamefont {L.~V.}\ \bibnamefont
  {Ahlfors}}\ and\ \bibinfo {author} {\bibfnamefont {L.}~\bibnamefont {Bers}},\
  }\href {https://doi.org/10.2307/1970141} {\bibfield  {journal} {\bibinfo
  {journal} {Annals of Mathematics}\ }\textbf {\bibinfo {volume} {72}},\
  \bibinfo {pages} {385} (\bibinfo {year} {1960})}\BibitemShut {NoStop}%
\bibitem [{\citenamefont {Bonk}(2006)}]{Bonk2006}%
  \BibitemOpen
  \bibfield  {author} {\bibinfo {author} {\bibfnamefont {M.}~\bibnamefont
  {Bonk}},\ }in\ \href {https://doi.org/10.4171/022-2/64} {\emph {\bibinfo
  {booktitle} {Proceedings of the International Congress of Mathematicians}}},\
  Vol.~\bibinfo {volume} {2}\ (\bibinfo  {publisher} {European Mathematical
  Society},\ \bibinfo {address} {Z{\"u}rich},\ \bibinfo {year} {2006})\ pp.\
  \bibinfo {pages} {1349--1373}\BibitemShut {NoStop}%
\bibitem [{\citenamefont {Astala}\ \emph {et~al.}(2005)\citenamefont {Astala},
  \citenamefont {P{\"a}iv{\"a}rinta},\ and\ \citenamefont
  {Lassas}}]{AstalaLassasPaivarinta2005}%
  \BibitemOpen
  \bibfield  {author} {\bibinfo {author} {\bibfnamefont {K.}~\bibnamefont
  {Astala}}, \bibinfo {author} {\bibfnamefont {L.}~\bibnamefont
  {P{\"a}iv{\"a}rinta}},\ and\ \bibinfo {author} {\bibfnamefont
  {M.}~\bibnamefont {Lassas}},\ }\href {https://doi.org/10.1081/PDE-200044485}
  {\bibfield  {journal} {\bibinfo  {journal} {Communications in Partial
  Differential Equations}\ }\textbf {\bibinfo {volume} {30}},\ \bibinfo {pages}
  {207} (\bibinfo {year} {2005})}\BibitemShut {NoStop}%
\bibitem [{\citenamefont {Li}\ and\ \citenamefont
  {Pendry}(2008)}]{LiPendry2008}%
  \BibitemOpen
  \bibfield  {author} {\bibinfo {author} {\bibfnamefont {J.}~\bibnamefont
  {Li}}\ and\ \bibinfo {author} {\bibfnamefont {J.~B.}\ \bibnamefont
  {Pendry}},\ }\href {https://doi.org/10.1103/PhysRevLett.101.203901}
  {\bibfield  {journal} {\bibinfo  {journal} {Physical Review Letters}\
  }\textbf {\bibinfo {volume} {101}},\ \bibinfo {pages} {203901} (\bibinfo
  {year} {2008})}\BibitemShut {NoStop}%
\bibitem [{\citenamefont {Fillion-Gourdeau}\ \emph {et~al.}(2021)\citenamefont
  {Fillion-Gourdeau}, \citenamefont {Lorin},\ and\ \citenamefont
  {MacLean}}]{FillionGourdeau2021}%
  \BibitemOpen
  \bibfield  {author} {\bibinfo {author} {\bibfnamefont {F.}~\bibnamefont
  {Fillion-Gourdeau}}, \bibinfo {author} {\bibfnamefont {E.}~\bibnamefont
  {Lorin}},\ and\ \bibinfo {author} {\bibfnamefont {S.}~\bibnamefont
  {MacLean}},\ }\href {https://doi.org/10.1103/PhysRevE.103.013312} {\bibfield
  {journal} {\bibinfo  {journal} {Physical Review E}\ }\textbf {\bibinfo
  {volume} {103}},\ \bibinfo {pages} {013312} (\bibinfo {year}
  {2021})}\BibitemShut {NoStop}%
\bibitem [{\citenamefont {Jones}\ and\ \citenamefont
  {Mahadevan}(2013)}]{JonesMahadevan2013}%
  \BibitemOpen
  \bibfield  {author} {\bibinfo {author} {\bibfnamefont {G.~W.}\ \bibnamefont
  {Jones}}\ and\ \bibinfo {author} {\bibfnamefont {L.}~\bibnamefont
  {Mahadevan}},\ }\href {https://doi.org/10.1098/rspa.2012.0653} {\bibfield
  {journal} {\bibinfo  {journal} {Proceedings of the Royal Society A:
  Mathematical, Physical and Engineering Sciences}\ }\textbf {\bibinfo {volume}
  {469}},\ \bibinfo {pages} {20120653} (\bibinfo {year} {2013})}\BibitemShut
  {NoStop}%
\bibitem [{\citenamefont {Cislo}\ \emph {et~al.}(2025)\citenamefont {Cislo},
  \citenamefont {Pavlopoulos},\ and\ \citenamefont
  {Shraiman}}]{CisloPavlopoulosShraiman2025}%
  \BibitemOpen
  \bibfield  {author} {\bibinfo {author} {\bibfnamefont {D.~J.}\ \bibnamefont
  {Cislo}}, \bibinfo {author} {\bibfnamefont {A.}~\bibnamefont {Pavlopoulos}},\
  and\ \bibinfo {author} {\bibfnamefont {B.~I.}\ \bibnamefont {Shraiman}},\
  }\href {https://doi.org/10.1103/PhysRevX.15.021068} {\bibfield  {journal}
  {\bibinfo  {journal} {Physical Review X}\ }\textbf {\bibinfo {volume} {15}},\
  \bibinfo {pages} {021068} (\bibinfo {year} {2025})}\BibitemShut {NoStop}%
\bibitem [{\citenamefont {Schimming}\ and\ \citenamefont
  {Beller}(2026)}]{schimming2026structure}%
  \BibitemOpen
  \bibfield  {author} {\bibinfo {author} {\bibfnamefont {C.~D.}\ \bibnamefont
  {Schimming}}\ and\ \bibinfo {author} {\bibfnamefont {D.~A.}\ \bibnamefont
  {Beller}},\ }\href@noop {} {\bibfield  {journal} {\bibinfo  {journal} {arXiv
  preprint arXiv:2610.27278}\ } (\bibinfo {year} {2026})}\BibitemShut {NoStop}%
\bibitem [{\citenamefont {Dzyaloshinskii}(1970)}]{dzyaloshinskii1970theory}%
  \BibitemOpen
  \bibfield  {author} {\bibinfo {author} {\bibfnamefont {I.}~\bibnamefont
  {Dzyaloshinskii}},\ }\href@noop {} {\bibfield  {journal} {\bibinfo  {journal}
  {Soviet Journal of Experimental and Theoretical Physics}\ }\textbf {\bibinfo
  {volume} {31}},\ \bibinfo {pages} {773} (\bibinfo {year} {1970})}\BibitemShut
  {NoStop}%
\bibitem [{\citenamefont {Anisimov}\ and\ \citenamefont
  {Dzyaloshinskii}(1972)}]{anisimov1973new}%
  \BibitemOpen
  \bibfield  {author} {\bibinfo {author} {\bibfnamefont {S.~I.}\ \bibnamefont
  {Anisimov}}\ and\ \bibinfo {author} {\bibfnamefont {I.~E.}\ \bibnamefont
  {Dzyaloshinskii}},\ }\href@noop {} {\bibfield  {journal} {\bibinfo  {journal}
  {Zh. Eksp. Teor. Fiz.}\ }\textbf {\bibinfo {volume} {63}},\ \bibinfo {pages}
  {1460} (\bibinfo {year} {1972})}\BibitemShut {NoStop}%
\bibitem [{\citenamefont {Hudson}\ and\ \citenamefont
  {Thomas}(1989)}]{hudson1989frank}%
  \BibitemOpen
  \bibfield  {author} {\bibinfo {author} {\bibfnamefont {S.~D.}\ \bibnamefont
  {Hudson}}\ and\ \bibinfo {author} {\bibfnamefont {E.~L.}\ \bibnamefont
  {Thomas}},\ }\href@noop {} {\bibfield  {journal} {\bibinfo  {journal}
  {Physical review letters}\ }\textbf {\bibinfo {volume} {62}},\ \bibinfo
  {pages} {1993} (\bibinfo {year} {1989})}\BibitemShut {NoStop}%
\bibitem [{\citenamefont {Vafa}\ \emph {et~al.}(2022)\citenamefont {Vafa},
  \citenamefont {Zhang},\ and\ \citenamefont {Nelson}}]{VafaZhangNelson2022}%
  \BibitemOpen
  \bibfield  {author} {\bibinfo {author} {\bibfnamefont {F.}~\bibnamefont
  {Vafa}}, \bibinfo {author} {\bibfnamefont {G.~H.}\ \bibnamefont {Zhang}},\
  and\ \bibinfo {author} {\bibfnamefont {D.~R.}\ \bibnamefont {Nelson}},\
  }\href {https://doi.org/10.1103/PhysRevE.106.024704} {\bibfield  {journal}
  {\bibinfo  {journal} {Physical Review E}\ }\textbf {\bibinfo {volume}
  {106}},\ \bibinfo {pages} {024704} (\bibinfo {year} {2022})}\BibitemShut
  {NoStop}%
\bibitem [{\citenamefont {Zhang}\ and\ \citenamefont
  {Nelson}(2022)}]{ZhangNelson2022}%
  \BibitemOpen
  \bibfield  {author} {\bibinfo {author} {\bibfnamefont {G.~H.}\ \bibnamefont
  {Zhang}}\ and\ \bibinfo {author} {\bibfnamefont {D.~R.}\ \bibnamefont
  {Nelson}},\ }\href {https://doi.org/10.1103/PhysRevE.105.054703} {\bibfield
  {journal} {\bibinfo  {journal} {Physical Review E}\ }\textbf {\bibinfo
  {volume} {105}},\ \bibinfo {pages} {054703} (\bibinfo {year}
  {2022})}\BibitemShut {NoStop}%
\bibitem [{\citenamefont {Driscoll}\ and\ \citenamefont
  {Trefethen}(2002)}]{driscoll2002schwarz}%
  \BibitemOpen
  \bibfield  {author} {\bibinfo {author} {\bibfnamefont {T.~A.}\ \bibnamefont
  {Driscoll}}\ and\ \bibinfo {author} {\bibfnamefont {L.~N.}\ \bibnamefont
  {Trefethen}},\ }\href@noop {} {\emph {\bibinfo {title} {Schwarz-christoffel
  mapping}}},\ Vol.~\bibinfo {volume} {8}\ (\bibinfo  {publisher} {Cambridge
  university press},\ \bibinfo {year} {2002})\BibitemShut {NoStop}%
\bibitem [{\citenamefont {Pollard}\ and\ \citenamefont
  {Morris}(2026)}]{PollardMorris2026}%
  \BibitemOpen
  \bibfield  {author} {\bibinfo {author} {\bibfnamefont {J.}~\bibnamefont
  {Pollard}}\ and\ \bibinfo {author} {\bibfnamefont {R.~G.}\ \bibnamefont
  {Morris}},\ }\href@noop {} {\bibinfo {title} {Defect kinematics in {2D}
  nematics: Contributions from surface topology, intrinsic and extrinsic
  geometry, solitons, defect orientations, and elastic anisotropy}} (\bibinfo
  {year} {2026}),\ \Eprint {https://arxiv.org/abs/2605.21855} {arXiv:2605.21855
  [cond-mat.soft]} \BibitemShut {NoStop}%
\bibitem [{\citenamefont {Myers}\ \emph {et~al.}(2024)\citenamefont {Myers},
  \citenamefont {Swift}, \citenamefont {R{\o}nning}, \citenamefont
  {Angheluta},\ and\ \citenamefont {Vi{\~n}als}}]{Myers2024}%
  \BibitemOpen
  \bibfield  {author} {\bibinfo {author} {\bibfnamefont {L.}~\bibnamefont
  {Myers}}, \bibinfo {author} {\bibfnamefont {C.}~\bibnamefont {Swift}},
  \bibinfo {author} {\bibfnamefont {J.}~\bibnamefont {R{\o}nning}}, \bibinfo
  {author} {\bibfnamefont {L.}~\bibnamefont {Angheluta}},\ and\ \bibinfo
  {author} {\bibfnamefont {J.}~\bibnamefont {Vi{\~n}als}},\ }\href
  {https://doi.org/10.1039/D3SM01616A} {\bibfield  {journal} {\bibinfo
  {journal} {Soft Matter}\ }\textbf {\bibinfo {volume} {20}},\ \bibinfo {pages}
  {2900} (\bibinfo {year} {2024})}\BibitemShut {NoStop}%
\bibitem [{\citenamefont {Mottram}\ and\ \citenamefont
  {Newton}(2014)}]{MottramNewton2014}%
  \BibitemOpen
  \bibfield  {author} {\bibinfo {author} {\bibfnamefont {N.~J.}\ \bibnamefont
  {Mottram}}\ and\ \bibinfo {author} {\bibfnamefont {C.~J.~P.}\ \bibnamefont
  {Newton}},\ }\bibfield  {journal} {\bibinfo  {journal} {arXiv preprint
  arXiv:1409.3542}\ }\href {https://doi.org/10.48550/arXiv.1409.3542}
  {10.48550/arXiv.1409.3542} (\bibinfo {year} {2014}),\ \Eprint
  {https://arxiv.org/abs/1409.3542} {arXiv:1409.3542 [cond-mat.soft]}
  \BibitemShut {NoStop}%
\bibitem [{\citenamefont {Joshi}\ \emph {et~al.}(2025)\citenamefont {Joshi},
  \citenamefont {Hellstein}, \citenamefont {Wennerholm}, \citenamefont
  {Downey}, \citenamefont {Hamilton}, \citenamefont {Hocking}, \citenamefont
  {Andrei}, \citenamefont {Adler},\ and\ \citenamefont
  {Atherton}}]{Joshi2025Morpho}%
  \BibitemOpen
  \bibfield  {author} {\bibinfo {author} {\bibfnamefont {C.}~\bibnamefont
  {Joshi}}, \bibinfo {author} {\bibfnamefont {D.}~\bibnamefont {Hellstein}},
  \bibinfo {author} {\bibfnamefont {C.}~\bibnamefont {Wennerholm}}, \bibinfo
  {author} {\bibfnamefont {E.}~\bibnamefont {Downey}}, \bibinfo {author}
  {\bibfnamefont {E.}~\bibnamefont {Hamilton}}, \bibinfo {author}
  {\bibfnamefont {S.}~\bibnamefont {Hocking}}, \bibinfo {author} {\bibfnamefont
  {A.~S.}\ \bibnamefont {Andrei}}, \bibinfo {author} {\bibfnamefont {J.~H.}\
  \bibnamefont {Adler}},\ and\ \bibinfo {author} {\bibfnamefont {T.~J.}\
  \bibnamefont {Atherton}},\ }\href
  {https://doi.org/10.1038/s43588-024-00749-7} {\bibfield  {journal} {\bibinfo
  {journal} {Nature Computational Science}\ }\textbf {\bibinfo {volume} {5}},\
  \bibinfo {pages} {170} (\bibinfo {year} {2025})}\BibitemShut {NoStop}%
\bibitem [{\citenamefont {Klein}\ and\ \citenamefont
  {Beller}(2026)}]{KleinBeller2026}%
  \BibitemOpen
  \bibfield  {author} {\bibinfo {author} {\bibfnamefont {B.}~\bibnamefont
  {Klein}}\ and\ \bibinfo {author} {\bibfnamefont {D.~A.}\ \bibnamefont
  {Beller}},\ }\href@noop {} {} (\bibinfo {year} {2026}),\ \bibinfo {note}
  {unpublished work}\BibitemShut {NoStop}%
\bibitem [{\citenamefont {Aharoni}\ \emph {et~al.}(2014)\citenamefont
  {Aharoni}, \citenamefont {Sharon},\ and\ \citenamefont
  {Kupferman}}]{Aharoni2014}%
  \BibitemOpen
  \bibfield  {author} {\bibinfo {author} {\bibfnamefont {H.}~\bibnamefont
  {Aharoni}}, \bibinfo {author} {\bibfnamefont {E.}~\bibnamefont {Sharon}},\
  and\ \bibinfo {author} {\bibfnamefont {R.}~\bibnamefont {Kupferman}},\ }\href
  {https://doi.org/10.1103/PhysRevLett.113.257801} {\bibfield  {journal}
  {\bibinfo  {journal} {Physical Review Letters}\ }\textbf {\bibinfo {volume}
  {113}},\ \bibinfo {pages} {257801} (\bibinfo {year} {2014})}\BibitemShut
  {NoStop}%
\bibitem [{\citenamefont {Modes}\ \emph {et~al.}(2010)\citenamefont {Modes},
  \citenamefont {Bhattacharya},\ and\ \citenamefont {Warner}}]{Modes2010}%
  \BibitemOpen
  \bibfield  {author} {\bibinfo {author} {\bibfnamefont {C.~D.}\ \bibnamefont
  {Modes}}, \bibinfo {author} {\bibfnamefont {K.}~\bibnamefont
  {Bhattacharya}},\ and\ \bibinfo {author} {\bibfnamefont {M.}~\bibnamefont
  {Warner}},\ }\href {https://doi.org/10.1103/PhysRevE.81.060701} {\bibfield
  {journal} {\bibinfo  {journal} {Physical Review E}\ }\textbf {\bibinfo
  {volume} {81}},\ \bibinfo {pages} {060701} (\bibinfo {year}
  {2010})}\BibitemShut {NoStop}%
\bibitem [{\citenamefont {Modes}\ \emph {et~al.}(2011)\citenamefont {Modes},
  \citenamefont {Bhattacharya},\ and\ \citenamefont
  {Warner}}]{Modes2011Gaussian}%
  \BibitemOpen
  \bibfield  {author} {\bibinfo {author} {\bibfnamefont {C.~D.}\ \bibnamefont
  {Modes}}, \bibinfo {author} {\bibfnamefont {K.}~\bibnamefont
  {Bhattacharya}},\ and\ \bibinfo {author} {\bibfnamefont {M.}~\bibnamefont
  {Warner}},\ }\href {https://doi.org/10.1098/rspa.2010.0352} {\bibfield
  {journal} {\bibinfo  {journal} {Proceedings of the Royal Society A:
  Mathematical, Physical and Engineering Sciences}\ }\textbf {\bibinfo {volume}
  {467}},\ \bibinfo {pages} {1121} (\bibinfo {year} {2011})}\BibitemShut
  {NoStop}%
\bibitem [{\citenamefont {Efrati}\ \emph {et~al.}(2009)\citenamefont {Efrati},
  \citenamefont {Sharon},\ and\ \citenamefont {Kupferman}}]{Efrati2009}%
  \BibitemOpen
  \bibfield  {author} {\bibinfo {author} {\bibfnamefont {E.}~\bibnamefont
  {Efrati}}, \bibinfo {author} {\bibfnamefont {E.}~\bibnamefont {Sharon}},\
  and\ \bibinfo {author} {\bibfnamefont {R.}~\bibnamefont {Kupferman}},\ }\href
  {https://doi.org/10.1016/j.jmps.2008.12.004} {\bibfield  {journal} {\bibinfo
  {journal} {Journal of the Mechanics and Physics of Solids}\ }\textbf
  {\bibinfo {volume} {57}},\ \bibinfo {pages} {762} (\bibinfo {year}
  {2009})}\BibitemShut {NoStop}%
\end{thebibliography}

\end{document}